\documentclass[preprint,12pt]{elsarticle}

\usepackage{amssymb}

\usepackage{graphicx}
\usepackage{subcaption}
\usepackage{caption}
\usepackage{multirow, multicol}

\usepackage{amsmath}
\usepackage{bm,amsmath}

\usepackage{lipsum}
\usepackage{blindtext}

\usepackage{array}
\newcolumntype{P}[1]{>{\centering\arraybackslash}p{#1}}
\usepackage[table,xcdraw]{xcolor}
\usepackage{multirow}

\usepackage{booktabs}

\usepackage{hyperref}

\usepackage{cleveref}

\begin{document}

\begin{frontmatter}



\title{Enhanced Speech Emotion Recognition with Efficient Channel Attention Guided Deep CNN-BiLSTM Framework}



\author[inst1]{Niloy Kumar Kundu}
\ead{nkundu191071@bscse.uiu.ac.bd}
\author[inst1]{Sarah Kobir}
\ead{skobir191214@bscse.uiu.ac.bd}
\author[inst1]{Md. Rayhan Ahmed\corref{cor1}}
\ead{rayhan@cse.uiu.ac.bd}
\author[inst1]{Tahmina Aktar}
\ead{taktar192060@bscse.uiu.ac.bd}
\author[inst1]{Niloya Roy}
\ead{nroy192057@bscse.uiu.ac.bd}

\cortext[cor1]{Corresponding author}

\affiliation[inst1]{organization={Department of Computer Science and Engineering, United International University}, 
            addressline={UIU Campus}, 
            city={Dhaka},
            postcode={1212}, 
            state={Dhaka},
            country={Bangladesh}}




\begin{abstract}
Speech emotion recognition (SER) is crucial for enhancing affective computing and enriching the domain of human-computer interaction. However, the main challenge in SER lies in selecting relevant feature representations from speech signals with lower computational costs. In this paper, we propose a lightweight SER architecture that integrates attention-based local feature blocks (ALFBs) to capture high-level relevant feature vectors from speech signals. We also incorporate a global feature block (GFB) technique to capture sequential, global information and long-term dependencies in speech signals. By aggregating attention-based local and global contextual feature vectors, our model effectively captures the internal correlation between salient features that reflect complex human emotional cues. To evaluate our approach, we extracted four types of spectral features from speech audio samples: mel-frequency cepstral coefficients, mel-spectrogram, root mean square value, and zero-crossing rate. Through a 5-fold cross-validation strategy, we tested the proposed method on five multi-lingual standard benchmark datasets: TESS, RAVDESS, BanglaSER, SUBESCO, and Emo-DB, and obtained a mean accuracy of 99.65\%, 94.88\%, 98.12\%, 97.94\%, and 97.19\% respectively. The results indicate that our model achieves state-of-the-art (SOTA) performance compared to most existing methods.
\end{abstract}


\end{frontmatter}

\section{Introduction}
\label{sec:Introduction}

The intricate process of human communication includes the exchange of knowledge between individuals using a variety of modalities, including speech, gestures, and facial expressions. Individuals regularly interact with information and emotions through speech \cite{el2011survey}. Speaking causes sound waves to be released into the environment, which the listener's ears pick up on. The listener's brain then decodes those sound waves to produce important details, such as the meaning of those words and their underlying emotions. Speech emotion recognition (SER) is an approach used to recognize and comprehend the emotional states of individuals by analyzing their speech, audio, and video. It works by extracting features from speech signals and mapping them to specific emotional states like happiness, sadness, anger, or fear \cite{sultana2021bangla}. SER provides various practical applications, such as improving interactions between humans and computers, otherwise known as human-computer interaction (HCI). Furthermore, in domains like psychology and healthcare, SER can be utilized to understand patients' emotional states and provide relevant interventions \cite{france2000acoustical}. Despite it's potential benefits, SER remains a challenging task. Due to the intricate nature of human emotions and the diverse characteristics of vocal signals, achieving accurate recognition results poses a significant challenge. \\

Feature extraction is a crucial part of building an accurate and robust SER model. The most essential and extensively used features in SER are acoustic features. In several studies, SER has been developed by incorporating a wide range of acoustic features, including prosodic features such as pitch, intonation, energy, and loudness, spectral features like centroid, flux, mel-frequency cepstral coefficients (MFCCs), and mel frequency magnitude coefficient, as well as voice-quality-related features such as jitter, shimmer, and harmonic-to-noise ratio \cite{n1,eyben2010towards, eyben2015geneva}. Additionally, features like the energy operator have also been incorporated into this process \cite{n1}. These features are used to acquire both the local and global contextual dependencies for the model to train upon in building SER systems. Both prosodic features and spectral features in speech contain emotional information, for that reason, they can be utilized to recognize speech emotions. Since prosodic and spectral features fluctuate over time, speech audio is segmented into frames of suitable sizes in SER systems. Low-level descriptor (LLD) features such as MFCC, Zero crossing rate (ZCR),  centroid, pitch, energy, Chromagram, root-mean-square (RMS), root-mean-square energy (RMSE), spectral contrast, and roll-off are widely used in the SER task \cite{ahmed2023ensemble}. In our study, we have extracted four different types of features from the speech audio and they are MFCC, mel-spectrogram, ZCR, and RMS value features. Selecting optimized features and an accurate classifier is a crucial aspect of SER because it can significantly impact the system's overall performance and preciousness. Over time, a wide variety of classifiers have been integrated for SER, and research in this field is ongoing. Some well-known classifiers in previous studies include Support Vector Machines (SVM) \cite{bhavan2019bagged}, Hidden Markov Model (HMM) \cite{8683172}, Gaussian Mixture Model (GMM) \cite{8651287}, K-nearest neighbors (KNN) \cite{venkata2021speech}, decision trees \cite{sun2019speech}, and ensemble approaches \cite{ahmed2023ensemble}. However, recent trends in SER research have shifted towards using deep learning-based frameworks such as Deep Neural Networks (DNN), Convolutional Neural Networks (CNN), and Recurrent Neural Networks (RNN) used in SER tasks \cite{8462677, ahmed2023ensemble, xu2022multi}. In addition, techniques like transfer learning, multitask learning, auto-encoders, attention mechanisms, adversarial training, etc, are also being utilized by researchers \cite{chen2023learning, lotfian2018predicting}. \\

\begin{figure}[htpb]
    \centering
    \includegraphics[width=\textwidth, height=6cm]{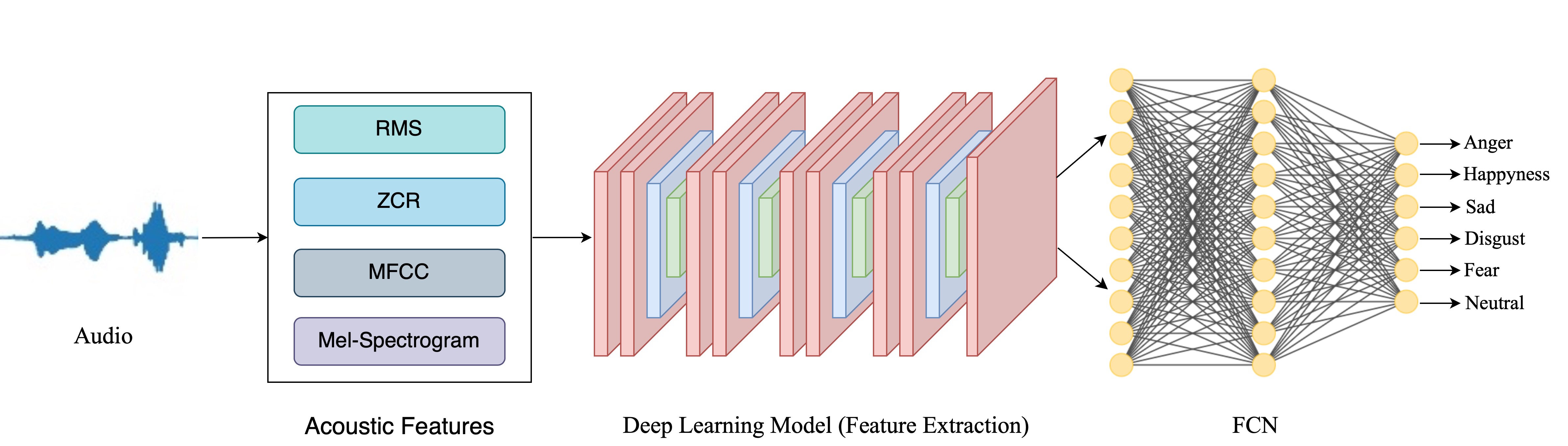}
    \caption{A graphical illustration of the proposed SER model's generic workflow.}
    \label{fig: basic workflow}
\end{figure}

Recent advancements in signal processing and deep learning-based methods have enriched the speech-processing domain. Deep learning is used for salient feature extraction which performs well for many image classification, SER, computer vision, and natural language processing-related tasks. Deep learning is essential in data science which has a close relationship to statistics and predictions \cite{lecun2015deep}. Researchers have discovered that deep learning-based methods can express speech features in higher dimensions, which can be highly challenging for traditional methods. Since each model is different from the other ones, there is bound to be some complementary information regarding the represented features. To fully understand and utilize these complementary aspects, numerous studies have been conducted in the field of deep learning-based speech feature combinations \cite{xu2022multi}. Studies showed that 1D CNN can extract the local features efficiently from sequential data \cite{chen2015convolutional}. The capability of 1D CNNs to extract these features from sequential data makes them highly useful across a wide range of applications. Time-series data are successfully processed using the 1D CNN model and audio classification tasks have shown tremendous promise \cite{vrysis2020experimenting}. To capture the long-term contextual correlations, temporal information, and understand the cues of emotions from speech, researchers have used Recurrent Neural Networks (RNN) \cite{graves2006connectionist}, Gated Recurrent Unit (GRU) \cite{cho2014learning, ahmed2023ensemble}, Stacked Ensemble, Long Short Term Memory (LSTM) \cite{hochreiter1997long}, Bidirectional Long Short-Term Memory (BiLSTM) \cite{schuster1997bidirectional} along with 1D CNN and Fully Connected Networks \cite{chen2021stock}. The CNN-based network can selectively focus on the most informative channels integrated with the channel-based attention method introduced by Wang et al. \cite{wang2020eca} which is an extremely lightweight efficient channel attention network (ECA-NET) module that calculates the relevance of each channel in the 1D CNN feature vectors. In recent years, ensemble learning  performed well in many sectors along with SER. Ensemble learning improves accuracy and robustness by combining the predictions of multiple models, decreasing the risk of over-fitting, and increasing the diversity of the learned representations using the 1D CNN model \cite{dietterich2002ensemble}. The main objective of incorporating multiple models in a single framework is to capture both spatial and temporal cues from the speech audios which can provide improved and robust SER performance. \\


Inspired by the notable achievements of deep learning-based models in the SER domain and the ability of attention mechanisms to re-calibrate features based on their relevance, we propose an architecture that integrates attention-based local feature blocks (ALFBs) which incorporates 1D CNN, ECA-Net, batch normalization, and 1D max-pooling for extracting high-level hidden local features. Parallel to the ALFBs the architecture also integrates global feature blocks (GFBs), which use  the BiLSTM network and batch normalization for extracting global features with contextual and temporal dependencies efficiently. This allows the proposed framework to acquire contextual dependencies and acquire necessary long-term patterns for the precise identification of emotions within the speech. To evaluate our approach, we employed five widely-used benchmark datasets of multiple languages from the public domain: Toronto Emotional Speech Set (TESS) \cite{dupuis2015aging}, Ryerson Audio-Visual Database of Emotional Speech and Song (RAVDESS) \cite{livingstone2018ryerson}, Bangla speech emotion recognition dataset (BanglaSER) \cite{das2022banglaser}, SUST Bangla Emotional Speech Corpus (SUBESCO) \cite{sultana2021sust}, and Berlin Database of Emotional Speech (Emo-DB) \cite{burkhardt2005database}. Furthermore, the domain of audio-based SER has seen limited exploration in the context of the Bangla language.

The limited number of samples present in individual classes in each of these datasets poses a significant challenge for the effective training of deep learning-based models. Additionally, some of these datasets have problems of class imbalance. To address these issues, we augmented the data, which improved our proposed model's generalization and overall performance. Figure \ref{fig: basic workflow}, represents a general and comprehensive high-level overview of the workflow of the proposed SER system. We extracted MFCC, mel-spectrogram, ZCR, and RMS value features from the audio samples as they are known to be beneficial for this task. The proposed model is trained using the mean values of these extracted features to enhance the recognition performance in detecting various human emotions, including ``happiness", ``sadness", ``surprise",  ``fearful", ``anger", ``boredom" and ``neutral" from audio signals. Our model demonstrates great state-of-the-art (SOTA) results for the SER task compared to most existing literature. This work's significant contributions are as follows:

\begin{itemize}
\item Our proposed model utilizes a dual channel architecture where in the first channel, we proposed four blocks of efficient channel attention-based local feature block (ALFB), and parallelly in the second channel, we incorporate two blocks of global feature block (GFB).

\item The ALFBs extract the hidden high-level local spatial features from the features extracted from speech audios by effectively combining 1D CNN and ECA-Net. It helps the model selectively focus on salient features and reduce the influence of less relevant features in the network.

\item The GFBs extract the contextual global features using BiLSTM which can further produce better feature representations to acquire more accurate emotional cues.

\item We have carried out a comprehensive set of experiments using five widely used, publicly accessible benchmark datasets incorporating English, German, and less-resourced Bengali language for SER: TESS, RAVDESS, BanglaSER, SUBESCO, and Emo-DB.

\item The performance of the proposed model is compared with the previous SOTA models. After conducting 5-fold cross-validation, our model achieves the SOTA mean accuracy of 99.65\% for TESS, 94.88\% for RAVDESS, 98.12\% for BanglaSER, 97.94\% for SUBESCO, and 97.19\% for Emo-DB  datasets. These are improved results compared to the previous SOTA methods on each dataset and provide great generalization ability.

\end{itemize}

The subsequent sections of this paper are organized as follows: Section II provides an overview of related works conducted in recent years, while Section III delves into the methodology and detailed discussion of the proposed framework. Section IV covers the datasets used and the features extracted in this study. The experimental analysis is presented in Section V, followed by a discussion of the obtained results and a comparison with other studies in Section VI. Lastly, Section VII concludes the paper with a summary of the findings.

\section{Related Works}
\label{sec: related works}

This section presents recent findings and studies related to SER tasks. \\
Deep learning is highly useful for data scientists who deal with collecting, analyzing, and understanding huge amounts of data. Deep learning-based models have performed tremendously well in SER systems. Kwon et al. \cite{A2} proposed a model based on a one-dimensional dilated CNN with residual blocks using a skip connection to recognize the local features from each segment of the speech signal. The proposed model was evaluated by the IEMOCAP and Emo-DB datasets with a prediction accuracy of 73\% and 90\% respectively. Alluhaidan et al. \cite{alluhaidan2023speech} combined both MFCCs and time-domain features (MFCCT) to enhance the SER performance. The proposed hybrid features were given to a CNN to build the SER model. The model achieves 97\%, 93\% and 92\% accuracy on the Emo-DB, SAVEE, and RAVDESS datasets, respectively. Patnaik \cite{patnaik2023speech} proposed a deep sequential model that classifies emotions based on complex mel frequency cepstral coefficients (c-MFCCs). The c-MFCCs capture both the magnitude and phase information of the speech signal which provides more comprehensive information. A 7-layered sequential deep CNN model is used to learn the temporal dependencies. Ahmed et al. \cite{ahmed2023ensemble} proposed a model that used MFCC, Chromagram and Pitch, LMS, ZCR, and RMS features to extract information from speech signals. They provided a novel approach for SER by combining 1D-CNN, LSTM, and GRU neural networks in an ensemble model. For the RAVDESS dataset, the suggested model achieved a weighted average accuracy of 95.62\%,  99.46\% on the TESS, 95.42\% on the Emo-DB, 93.22\% on  SAVEE and 90.47\% on CREMA-D datasets, respectively. Zhong \cite{zhong2023speech} presents an SER system based on SVM and CNN with MFCC feature extraction. The author obtained two accuracies that are low. The penalty coefficient and Gamma value have been changed to improve the accuracy of SVM. For CNN, the dropout layer has been added to the CNN structure and changed the L2 normalizer and number of epochs to increase the accuracy. The author used the CASIA Chinese emotional speech database to train the model. Sadia et al. \cite{sultana2021bangla} proposed a model which used deep CNNs and a BiLSTM with a time-distributed flatten layer for the SER focusing on the Bangla language. The proposed model has attained a SOTA perceptual efficiency achieving weighted accuracies of 86.9\%, and 82.7\% for the SUBESCO and RAVDESS datasets, respectively. Aggarwal et al. \cite{A13} provided two different feature extraction methods to address successful SER. The first set of features has been obtained using principal component analysis (PCA) along with a DNN with dense and dropout layers. In the second approach, they extracted mel-spectrogram images from the audio files. After that, the images are sent as input to the pre-trained Visual Geometry Group-16 (VGG-16) \cite{simonyan2014very} model. They tested their model using two datasets. The accuracy of the proposed-I model was 73.95\% and 99.99\% on the RAVDESS and TESS datasets, respectively. On the other hand, the accuracy of the proposed-II model was 81.94\% and 97.15\% on the RAVDESS and TESS datasets, respectively. Ala Saleh Alluhaidan et al. \cite{alluhaidan2023speech} proposed a hybrid feature called MFCCT which is a combination of MFCC and time-domain features. The derived features were used as input for an 1D CNN model. On the Emo-DB, SAVEE, and RAVDESS datasets, using hybrid MFCCT feature and 1D CNN achieved an accuracy of 97\%, 93\%, and 92\%, respectively. \\

Nowadays, attention mechanisms are widely known to have a lot of potentials to improve the performance of deep CNNs. Some attention architectures that have been widely used by researchers include Channel Attention, Squeeze-and-Excitation Networks (SEnet) \cite{hu2018squeeze}, Convolutional Block Attention module (CBAM) \cite{woo2018cbam}, and ECA-Net \cite{wang2020eca}, etc. Zou et al. \cite{A3} proposed a model using three different encoders CNN, BiLSTM and the transformer-based wav2vec2. These encoders help to extract the acoustic information. CNN is used for extracting the spectrogram. Similarly, for MFCC, they used BiLSTM, and for extracting the raw audio signals wav2vec2 is also used. In the wav2vec2 embedding, they applied a co-attention module. Based on the MFCC and spectrogram features, the attention module assigns weights to each frame. After that, they combine all three extracted features. The experimental results achieved an weighted accuracy of 71.64\% and an unweighted accuracy of 72.70\% on the IEMOCAP dataset. Zhao et al. \cite{zhao2021combining} proposed a hybrid architecture that uses parallel convolutional layers integrated with SEnet to extract relationships from 3D spectrograms. In the classification block, they introduced a self-attention Residual Dilated Network (SADRN) with Connectionist Temporal Classification (CTC) loss for discrete SER. The suggested technique achieves a weighted accuracy of 73.1\%, unweighted accuracy of 66.3\% on IEMOCAP, and an unweighted accuracy of 41.1\% on the FAU-AEC dataset, respectively. Mustaqeem et al. \cite{mustaqeem2021speech} proposed a self-attention module that receives a transitional feature map. It produces a channel and spatial attention map. They incorporate dilated CNN in spatial attention for extracting spatial information. Moreover, a multi-layer perceptron in channel attention is also added to extract global features. The accuracy of the proposed model is 78.01\%, 80.00\%, and 93.00\% over IEMOCAP, RAVDESS, and EMO-DB datasets, respectively. Sun et al. \cite{A9} developed a multimodal cross and self-attention network. The cross-attention module captures inter-modal interactions between acoustic frames and textual words in pairs. Meanwhile, the self-attention module utilizes the self-attention mechanism to propagate information within each modality. The model achieved a weighted accuracy of 61.2\% and an unweighted accuracy of 56\% over the IEMOCAP dataset. For the MELD dataset, the model achieved an F1-score of 59.2\%. \\ 

Ensemble learning is used for improving predicting performance by combining multiple models. Researchers use ensemble learning to increase ML models' accuracy, robustness, and generalization performance. Zhang et al. \cite{zhang2021speech} proposed an ensemble model that combines a random forest classifier with the weighted binary cuckoo search method to select the optimal feature subset. Falahzadeh et al. \cite{falahzadeh2023deep} proposed a new transform of speech signals into a chromatogram which is derived from the reconstructed phase space of speech. This chromatogram is basically a color image. The extracted images are passed through the DCNN with a gray wolf optimization module for optimizing the hyperparameters of the model. Chalapathi et al. \cite{chalapathi2022ensemble} proposed a model that utilizes the adaptive boosting ensemble method and the fuzzy c-means approach. This model extracts high-dimensional acoustic features for emotion recognition from speech audio signals using the benchmark dataset RAVDESS. Zvarevashe et al. \cite{A11} proposed a stacked ensemble algorithm to recognize a cross-lingual acoustic emotional valence by using logistic regression, random decision forest, gradient boosting, and AdaBoost algorithms. Five speech emotion corpora SAVEE, EMO-DB, RAVDESS, CREMA-D, and EMOVO that serve as the study's materials are combined in the proposed models. 

In contrast to the studies discussed above, the primary objective of this paper is to exploit the distinctive attributes of diverse neural network architectures. This is done to effectively capture both low-level and high-level representations of speech features within an optimized attention-guided framework, which emphasizes the preservation of salient feature vectors. An extensive comparative evaluation is provided in \Cref{table2,table3,table4,table5,table6}. This comparison is performed between the notable works discussed in the literature review section and our proposed work. In the comparison, we highlight utilized datasets, the methodology, extracted features, and achieved results.

\section{Methodology}
\label{sec: Proposed Methods}
\begin{figure}[h]
    \centering
    \includegraphics[width=\textwidth, height=0.7\textwidth]{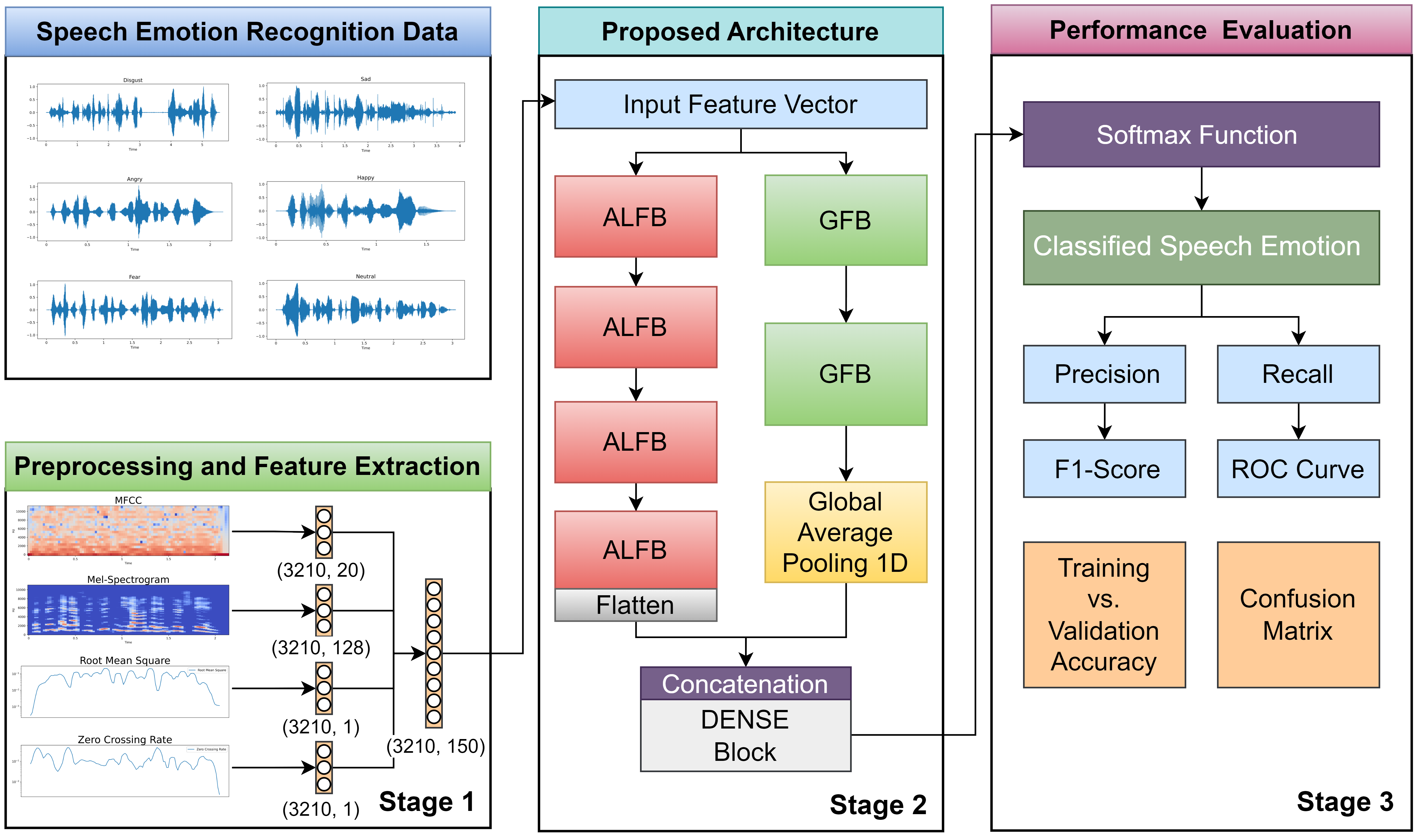}
    \caption{Overview of the proposed approach for SER.}
    \label{fig:proposedApproach}
\end{figure}

As depicted in Figure \ref{fig:proposedApproach}, our initial step involved utilizing the SER dataset. In Stage 1, we extracted MFCC, mel-spectrogram, RMS, and ZCR values, which were then represented as a feature vector. This feature vector was combined and the dimension is (3210, 150). Subsequently, we split the feature vector into training, validation, and testing sets with a ratio of 60:25:15 respectively.
For model training, we fed both the training and validation data. Moving on to Stage 2, our proposed model employed four ALFBs and two GFBs to extract local and global features. Then we concatenated both ALFB and GFB. The classification results were obtained by applying the softmax function after performing the execution of the dense block. Finally, we evaluated our model using the test data. Additionally, we conducted 5-fold cross-validation, which will be discussed in the upcoming sections. The evaluation metrics are presented in Stage 3.

\begin{figure}[ht!]
    \centering
    \includegraphics[width=0.9\textwidth, height=1.1\textwidth]{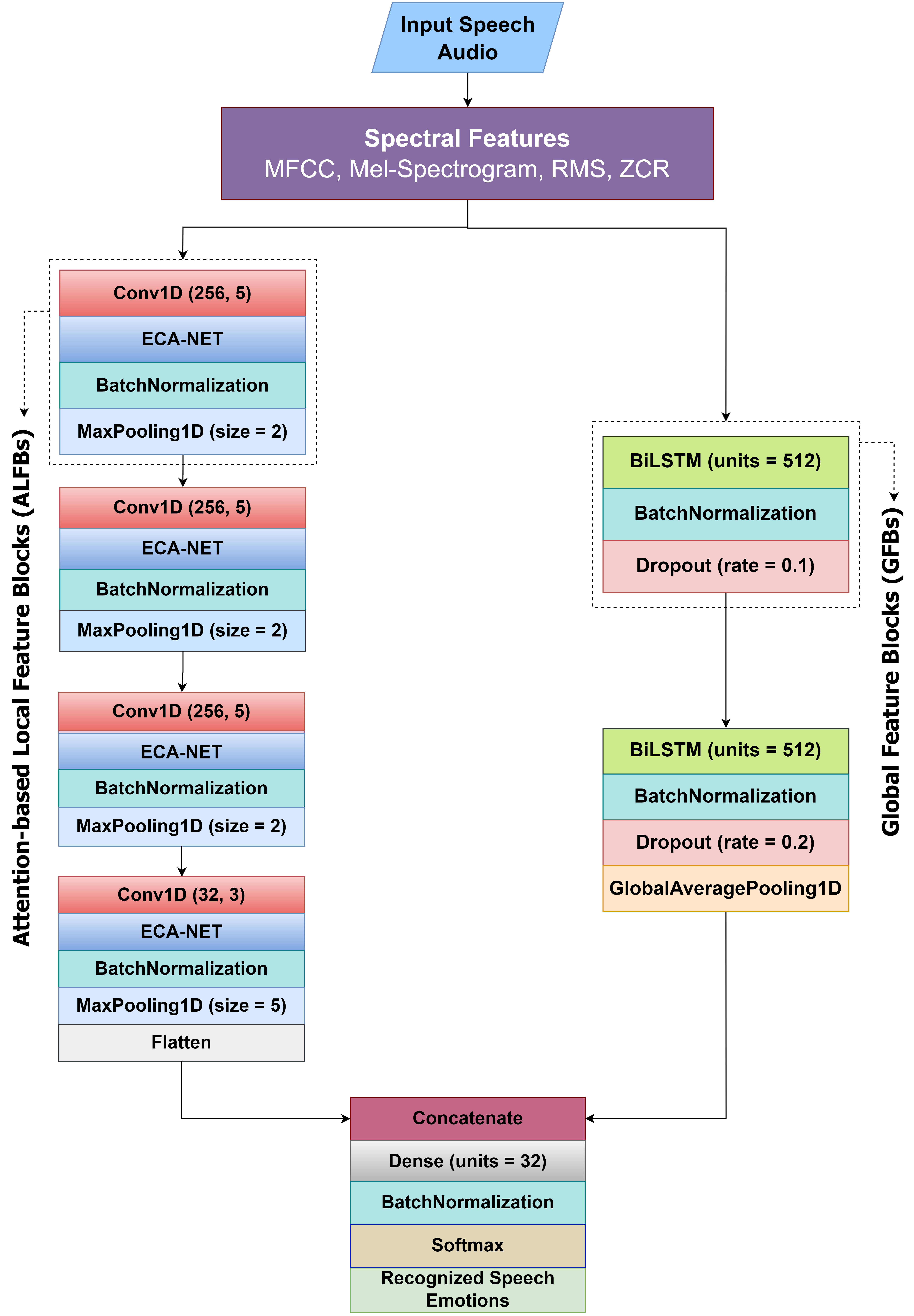}
    \caption{Visual representation of the proposed dual-channel architecture.}
    \label{fig:model_summary}
\end{figure}

This section provides a detailed overview of the proposed architecture as depicted in Figure \ref{fig:model_summary}.

\subsection{Proposed Dual Channel SER Model}

We have designed dual-channel attention-guided 1D CNN-BiLSTM networks-based deep neural networks. One channel is used for extracting high-level hidden local features. The other channel is utilized for contextual and sequential global feature extraction. In addition, ECA-Net employs a parallel convolutional structure that enables it to capture both local and global contextual information in the input signal, which can be important for recognizing complex emotional expressions.

\subsubsection{Attention Based Local Feature Blocks}

The channel for extracting local features is made of four ALFBs. Each block has four distinct sections. The first part is the 1D CNN layer which contains 256 filters, a kernel size of five, padding = `same', strides = 1, and the Rectified Linear Unit (ReLU) is used as an activation function. ReLU is used for solving the vanishing gradient problem in our model. The 1D CNN is effective in extracting local features from sequential data. It can learn spatial and temporal patterns in the data, capturing short-term patterns and structures that are beneficial for emotion recognition. Then the output of the 1D CNN layer is passed through the second part an ECA-Net block, which is responsible for channel attention.

\begin{figure}[htpb]
    \centering
    \includegraphics[width=0.9\textwidth, height=0.3\textwidth]{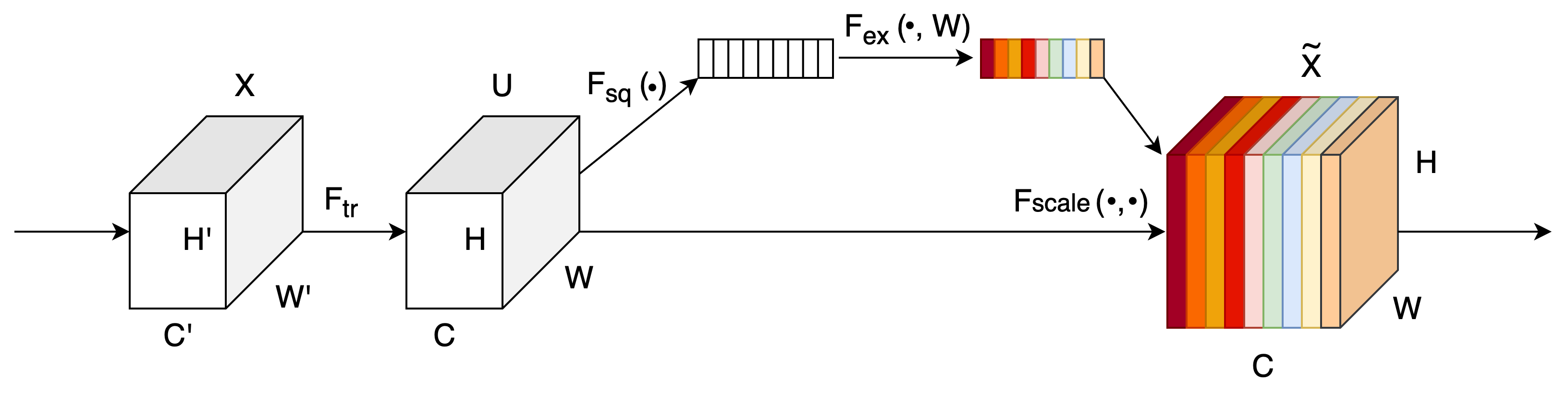}
    \caption{Visual representation of the Squeeze and Excitation Network block \cite{hu2018squeeze}.}
    \label{fig:se-net}
\end{figure}

ECA-Net is a lightweight attention-mechanism module derived from Squeeze and Excitation Network (SE-Net). The SE-Net is one of the representative models of the channel attention mechanism. It was the champion model of the final ImageNet classification competition in 2017. It performs three main operations: squeeze, excitation, and reweight. Figure \ref{fig:se-net} shows the SE-Net module.

In figure \ref{fig:se-net}, $X$ is the Input feature map, where C, H, and W are Channel dimensions, height, and width, respectively. To describe the outcome of a given transformation, \(F_{\text{tr}}: X \rightarrow U, X \in R^{H' \times W' \times C'}, U \in R^{H \times W \times C}\) and $\Tilde{X}$ represent the resulting feature map. $F_{sq}$ represents the squeeze operation that reduces the spatial dimensions of each feature map to 1x1, resulting in a channel-wise representation. The calculation formula is given below:

\begin{equation}
    \label{eq: channel descriptor}
    \begin{split}
        \text{$F_{sq}(u_c)$}&=\frac{1}{H \times W} \sum_{i=1, j=1}^{H, W} u_c(i,j)
    \end{split}
\end{equation}

$F_{ex}$ denotes the excitation operation. The goal here is to learn a set of weights for each channel of the input feature map $X$ to capture the importance of each channel. These weights are learned using a small neural network consisting of one or more fully connected layers. The calculation formula is:

\begin{equation}
    \label{eq:excitation}
    \begin{split}
        F_{\text{ex}(z, W)} = \sigma(g(z, W)) = \sigma(W_2 \delta(W_1z))
    \end{split}
\end{equation}

In Eq \ref{eq:excitation}, W1 and W2 represent the two fully connected layers. $F_scale$ denotes the scaling operation which is also known as reweighting. This is done by multiplying each channel of the original feature maps. For this reason, it gives more importance to the channels that are important and reduces the significance of less important channels. The result is a feature map that is focused on the most relevant information, which can be beneficial for improving the performance of deep neural networks.

\begin{figure}[htpb]
    \centering
    \includegraphics[width=0.9\textwidth, height=0.5\textwidth]{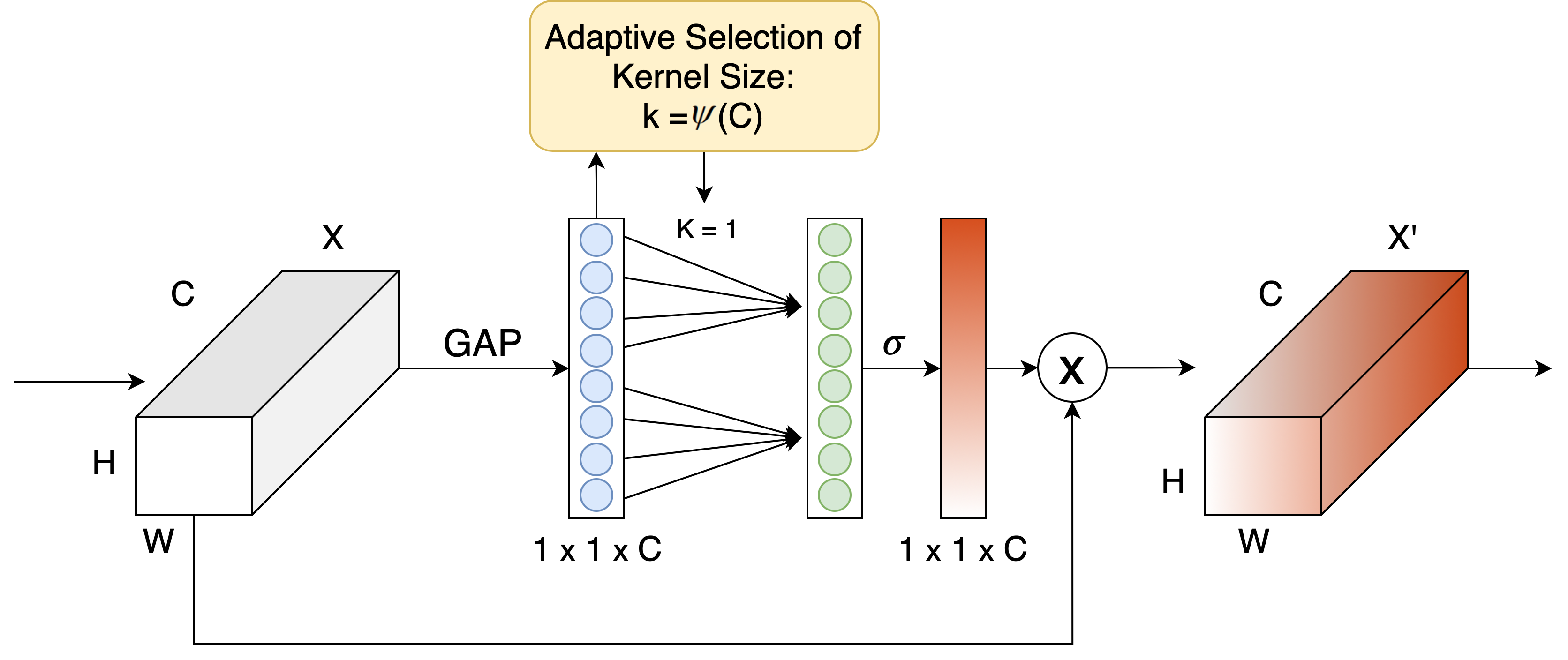}
    \caption{Visual representation of the Efficient Channel Attention Network block \cite{wang2020eca}.}
    \label{fig:eca-net}
\end{figure}

However, SE-Net also has some disadvantages. Sometimes, it loses important information during the process of squeezing and capturing the dependencies of all channels. To overcome this issue, ECA-Net is proposed. It allows the network to selectively focus on the most informative features. For that reason, it can easily suppress the noise and irrelevant information. Figure \ref{fig:eca-net} represents the diagram of the ECA-Net module.

In Figure \ref{fig:eca-net}, X is the Input feature map, where C, H, and W are Channel dimensions (number of filters), height, and width, respectively. A global average pooling (GAP) operation is applied to the input feature map X which compresses $H \times W \times C$ data into $1 \times 1 \times C$ without dimensionality reduction and the features of each channel are aggregated. In ECA-Net, first, we adaptively select the kernel size $K = 1$ by a function of channel dimension C. After that, a one-dimensional convolution of size $K$ is performed. Finally, a sigmoid function has been performed to learn channel attention. The convolution kernel size ($K$), plays an important role in identifying the coverage of interaction within the data. This parameter is related to the channel dimension C, which is typically chosen to be a power of 2. The mapping relationship between these two parameters can be calculated as follows:

\begin{equation}
    \label{eq:excitation}
    \begin{split}
        C = \phi(k) = 2^{(\gamma \times k - b)}
    \end{split}
\end{equation}

The computational formula of the convolution kernel size $K$ is given below. 

\begin{equation}
    \label{eq:excitation}
    \begin{split}
        k = \psi(C) = \left|\frac{\log_2(C)}{\gamma} + \frac{b}{\gamma}\right|_{\text{odd}}
    \end{split}
\end{equation}

Here, $|t|_{odd}$ indicates the nearest odd number of $t$, $\gamma$ is set to 2, and $b$ to 1.

After that, the output feature map from ECA-Net is reshaped so that it can pass through the third part of our model which is batch normalization (BN). It increased the training stability and speed of the network. As a result, it is passed through the final part 1D max-pooling layer with a pool size of two, a stride of one, and using the padding as `same'. It is used to produce the same output length similar to the input sequence length while keeping the most important features. The following three ALFBs are employed, with kernel sizes of 5, 5, and 3 for the 1D convolution layers. The filters on these blocks are 256 except for the last block which is 32. Padding and stride configurations are the same as the previous blocks. In addition, ECA-Net is also performed after each 1D convolution layer. Batch normalization is also passed through these layers and a 1D max-pooling with the following three ALFBs with a pool size of 2, 2, 5, and a stride of two is performed. After that, we passed this input into a flattened layer.

\subsubsection{Global Feature Blocks}
The global feature blocks (GFBs) are responsible for global contextual feature extraction and consist of BiLSTM. It can capture how speech features change over time by processing speech frames or signals in both forward and backward directions. They take into account data from previous and forthcoming frames, allowing the model to comprehend the larger context and identify long-term patterns crucial for precise emotion recognition in speech. For the SER task, it is important because emotional cues in speech often occur over a longer period of time than just a few milliseconds. For that reason, it is influenced by the context of preceding and succeeding speech segments.

\begin{figure}[htpb]
    \centering
    \includegraphics[width=0.9\textwidth]{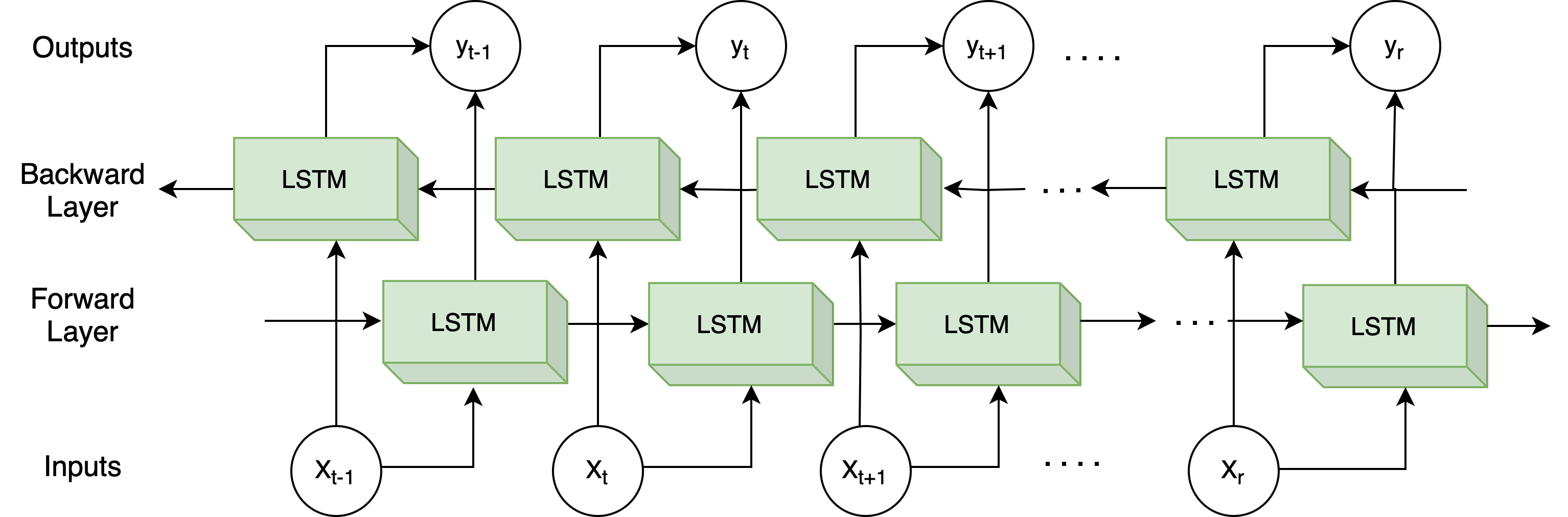}
    \caption{Visual representation of the Bidirectional Long Short Term Memory (BiLSTM).}
    \label{fig:BiLSTM}
\end{figure}

The LSTM unit has four fundamental components, including one input gate $i_t$ with corresponding weight matrices $W_{x_i}$, $W_{c_i}$, $W_{h_i}$ and bias $b_i$, one forget gate $f_t$ with corresponding weight matrices $W_{x_f}$, $W_{c_f}$, $W_{h_f}$ and bias $b_f$, one output gate $o_t$ with corresponding weight matrices $W_{x_o}$, $W_{c_o}$, $W_{h_o}$ and bias $b_o$, and the cell state $c_t$. Each of these gates calculates its output based on the current input $x_i$, the previous hidden state $h_{(t-1)}$, and the previous cell state $c_{(t-1)}$. Simultaneously, the same input sequence is passed through the backward LSTM layer. It processes the sequence in reverse order. Additionally, it also maintains hidden states and memory cell states. The computational formula is given below:

\begin{equation}
i_t = \sigma(W_{x_i} x_t + W_{h_i}h_{(t-1)} + W_{c_i}c_{(t-1)} + b_{i})
\end{equation}

\begin{equation}
f_t = \sigma(W_{x_f}x_t + W_{h_f}h_{(t-1)} + W_{c_f}c_{(t-1)} + b_{f})
\end{equation}

\begin{equation}
g_t = \tanh(W_{x_c}x_t + W_{h_c}h_{(t-1)} + W_{c_c}c_{(t-1)} + b_{c})
\end{equation}

\begin{equation}
o_{t} = \sigma(W_{x_o}x_{t} + W_{h_o}h_{(t-1)} + W_{c_o}c_t + b_{o})
\end{equation}

\begin{equation}
c_t = f_t \odot c_{(t-1)} + i_t \odot g_t
\end{equation}

\begin{equation}
h_{t} = o_{t} \odot \tanh(c_{t})
\end{equation}

The final output of the BiLSTM is calculated by concatenating the forward and backward hidden states at each time step. Eq. \ref{Eq: hidden add} represents the final output.

\begin{equation}
h_t = [\overrightarrow{h_t} \oplus \overleftarrow{h_t}]
\label{Eq: hidden add}
\end{equation}

In our proposed model, BiLSTM has been performed using the input from the input layer with units of 512 and making the return sequence true. Then it is passed through a BN layer to improve the training stability and efficiency of the network. Furthermore, we have used dropout with a dropout rate of 0.1 to prevent over-fitting and improve the generalization performance of our model. The next GFB used the same configuration as the previous GFB except for the dropout rate. In this block, we incorporate the dropout rate is 0.2 in the dropout layer. After that, the output of the dropout layer is passed through the 1D global average pooling layer and merged with the output of the other channel.

Finally, we concatenated both the final ALFB and GFB together to enhance the model's performance with better feature representation consisting of hidden local emotional cues as well as long-term contextual relations and dependencies. After merging both the local and global features, it is passed through a dense layer with a unit of 32, followed by a batch normalization layer. The output layer utilizes the softmax activation function to differentiate between the emotions present in the speech audio.

\section{Dataset and Feature Extraction}

\subsection{Dataset}

For this study, we have used five different datasets: TESS, RAVDESS, BanglaSER, SUBESCO, and Emo-DB which cover English, Bengali, and German languages.

\textbf{TESS}: The TESS \cite{dupuis2015aging} dataset is a collection of high-precision, high-cadence time-series photometry data gathered by the TESS mission. This dataset contains 200 target words that are spoken by two actresses, one aged 26 and the other aged 64. The dataset comprises 2800 audio recordings and is nicely balanced and represents seven emotions: ``happy", ``neutral", ``angry", ``disgust", ``fear", ``surprise", and ``sad".

\textbf{RAVDESS}: The RAVDESS \cite{livingstone2018ryerson} has been an extensively used dataset in the field of SER. It contains audio and video recordings of 12 male and 12 female performers. This dataset contains 8 different emotional expressions such as ``sad", ``fearful",  ``happy", ``calm", ``angry", ``surprised", ``disgust", and ``neutral". In this research, our focus was on utilizing speech audio samples. There are 1440 audio samples overall. Each audio sample has a sample rate of 48 kHz and 60 trials per performer. It is a balanced dataset except for the ``neutral" class. It contains fewer records than the other classes.

\textbf{BanglaSER}: The BanglaSER \cite{das2022banglaser} is an important resource for research on SER in Bengali, a language spoken by over 200 million people. The banglaSER dataset comprises speech-audio data from 34 participating speakers. It includes an equal distribution of 17 male and 17 female nonprofessional actors from diverse age groups ranging between 19 and 47 years. The dataset has a sampling rate of 44.1 kHz with 306 recordings for each of ``happy", ``sad", ``angry", and ``surprised" emotions, but there were 243 recordings available for the ``neutral" emotion.

\textbf{SUBESCO}: The SUBESCO \cite{sultana2021sust} dataset has been used in ML and deep learning studies on emotion recognition in speech. This dataset is a gender-balanced dataset. The dataset includes voice data from 20 professional speakers, with an equal distribution of 10 male and 10 female speakers. Each of them expressed seventh different emotions: ``sadness", ``happiness", ``surprise", ``anger", ``disgust", ``fear", and ``neutral". In the context of the Bengali language, this dataset stands as the largest with a total duration of 7 hours, and encompasses 7000 utterances.

\textbf{Emo-DB}: The Emo-DB \cite{burkhardt2005database} dataset is a collection of emotional audio clips in the German language, consisting of 535 recordings from 10 German speakers (5 male and 5 female) expressing seven distinct emotions: ``anger", ``boredom", ``disgust", ``sadness", ``happiness", ``fear", and ``neutral". Each recording is accompanied by subtitles indicating the speaker's gender, the spoken sentence, and the expressed emotional state.

\subsection{Data Augmentation}

In the field of SER, data augmentation serves as a valuable technique for expanding the quantity and diversity of training data, which can enhance the effectiveness and generalization of ML-based models. By performing various transformations on the existing data, data augmentation involves producing new training samples. Due to the relatively low number of speech utterance records in each class, this study employs five types of audio data augmentation techniques., Additive White Gaussian Noise (AWGN) injection \cite{awgn}, pitch shifting, time-stretching, adding noise on the pitch shifting data, and adding pitch shifting on the time-stretching data. We added AWGN to the samples by using NumPy's uniform method with a rate of 0.015. Pitch shift was performed using the librosa library in Python by employing a factor of 0.7. Speed or Duration has been added without changing the pitch by stretching the time using the method called time\_stretch - python’s librosa library with a factor of 0.8. We also added noise to the pitch-shifted audio files. In addition, we also shifted the pitch of the audio after stretching the time using the same procedure. Notably, these augmentation techniques were implemented without compromising the performance of the SER system. After performing data augmentations, the overall samples of TESS, RAVDESS, BanglaSER, SUBESCO, and Emo-DB datasets increased to 16799, 8637, 8796, 41991, and 3210, respectively. \\

\subsection{Extracted Features}
To enhance the performance of our model, we have extracted various spectral features as the initial feature set which include MFCC, mel-spectrogram, ZCR, and RMS values from the audio samples.
MFCC is one of the widely used features for the SER task \cite{ahmed2023ensemble, patnaik2023speech}. We focused on MFCCs because they are often used in speech processing and are good candidates for speech feature extraction since they capture the spectrum features of audio signals. They are particularly useful for conveying phonetic and prosodic information because they closely resemble the way the human auditory system reacts to sound. They can aid in capturing variations in speech that express emotion, like shifts in pitch, intonation, and timbre, in the process of recognizing emotions \cite{dolka2021speech}. The extraction process relies on the inherent characteristics of the human auditory system, which serves as a natural reference for speech recognition. To obtain MFCC, the speech frame undergoes an initial application of a hamming window. Subsequently, the discrete spectrum is computed using a discrete fourier transform (DFT) \cite{liu2018speech}. MFCC can be calculated through Eq. \ref{eq: MFCC}, where $N$ is the sample length, $h(n)$ is the hamming window and $K$ is the length of the DFT.

\begin{equation}
    \label{eq: MFCC}
    \begin{split}
        \text{$y_i(k)$}&= \sum_{n = 1}^{N} y_i(n)h(n)e^{\frac{-j2\pi kn}{N}} \quad {;} \quad {1 \le k \le N}
    \end{split}
\end{equation}

The human auditory system perceives frequencies on a logarithmic scale, and the mel-spectrogram is a logarithmic representation of signal frequencies that share this characteristic. Mel-spectrogram visualizes the power of frequency bands over time, showing the relative importance of different frequency bands, similar to how our ears perceive sound \cite{leitner2019audio}. We use the mel-spectrogram to depict the spectrum features of speech. With MFCCs, it is a helpful function that is commonly used. Mel-spectrograms can highlight important frequency components and acoustic patterns in audio that point to emotional content. Mel-spectrogram helps to identify important information related to the spectral content and the frequency distribution of speech \cite{lu2020speech}. The correlation between the mel spectrum frequency ($f_{mel}$) and the signal frequency ($f_{Hz}$) can be characterized as follows:

\begin{equation}
    \label{eq: ZCR}
    \begin{split}
        \text{$f_{mel}$}&= 2595 .\log_{10} (1 + \frac{f_{Hz}}{700})
    \end{split}
\end{equation}

ZCR records the frequency at which the audio stream crosses the threshold of zero amplitude. We looked at it because of the information it can provide about how rapidly the audio signal changes and how it relates to speech tempo and prosody. ZCR shifts during emotive speech may be a reflection of rhythmic changes that might convey subtle feelings  \cite{koduru2020feature}. In addition, ZCR divides the number of samples in a specific region of an audio frame. After that, it provides the number of zero crossings in that region. Mathematically, the ZCR can be calculated using Eq. \ref{eq: ZCR}, where x represents the length of N. For the positive sample amplitude, the $\alpha$ function provides 1 and 0 for the negative sample amplitude over a time frame (t).

\begin{equation}
    \label{eq: ZCR}
    \begin{split}
        \text{$ZCR_t$}&= \frac{1}{N - 1} \sum_{i=1}^{N-1} \alpha(x_i x_{i-1})
    \end{split}
\end{equation}

\[ \text{where}, \alpha =  \begin{cases}
      1, & x_i x_{i-1} \geq 0 \\
      0, &  x_i x_{i-1} \le 0 \\
    \end{cases}
\]

RMS is used to measure the loudness of the sound. RMS is widely used by researchers for the SER task that uses the magnitude of the audio signal \cite{ahmed2023ensemble, er2020novel}. It provides information about the intensity of the speech signals. It measures the average power of an audio signal. It is practical for spotting variations in loudness and intensity, which may be a sign of various emotional states \cite{bhangale2023speech}. RMS can be calculated by considering the square root of the sum of the mean squares of the amplitudes of the sound samples. Eq. \ref{eq: RMS} represents the formula of RMS.

\begin{equation}
    \label{eq: RMS}
    \begin{split}
        \text{$x_{rms}$}&= \sqrt{ \frac{1}{N} \sum_{i=1}^{N} x_i^2}
    \end{split}
\end{equation}

In this study, we extracted a total of 20 MFCC features, 128 Mel-spectrogram features, as well as one RMS and one ZCR feature. These features are combined to create a feature vector with a dimension of $3210 \times 150$.

\section{Experimental Analysis}
In this section, we discuss about the Experimental Setup, and Hyperparameter Tuning for the model's optimization.
\label{sec: Experimental Analysis}


\subsection{Experimental Setup}

\begin{table}[htbp]
\centering
\caption{Structure of the proposed architecture.}
\label{table: structure of propose model}
\resizebox{0.9\textwidth}{!}{%
\begin{tabular}{@{}cclc@{}}
\toprule
\multicolumn{1}{l}{Block Name} & Block No.                  & Detail                                           & \multicolumn{1}{l}{Parameters} \\ \midrule
\multirow{17}{*}{ALFBs} &
  \multirow{4}{*}{Block - 1} &
  Conv1D: 256 filters, 5 × 5 kernels, 1 × 1 stride &
  \multirow{4}{*}{2562} \\
                               &                            & ECA-Net Module                                   &                                \\
                               &                            & Batch Normalization                              &                                \\
                               &                            & MaxPool1D: 2 × 2 pool size, 2 × 2 stride         &                                \\

    \cline{2-4}
                               
                               & \multirow{4}{*}{Block - 2} & Conv1D: 256 filters, 5 × 5 kernels, 1 × 1 stride & \multirow{4}{*}{328962}        \\
                               &                            & ECA-Net Module                                   &                                \\
                               &                            & Batch Normalization                              &                                \\
                               &                            & MaxPool1D: 2 × 2 pool size, 2 × 2 stride         &                                \\

    \cline{2-4}
                               
                               & \multirow{4}{*}{Block - 3} & Conv1D: 256 filters, 5 × 5 kernels, 1 × 1 stride & \multirow{4}{*}{328962}        \\
                               &                            & ECA-Net Module                                   &                                \\
                               &                            & Batch Normalization                              &                                \\
                               &                            & MaxPool1D: 2 × 2 pool size, 2 × 2 stride         &                                \\

    \cline{2-4}
    
                               & \multirow{4}{*}{Block - 4} & Conv1D: 32 filters, 3 × 3 kernels, 1 × 1 stride  & \multirow{4}{*}{24738}         \\
                               &                            & ECA-Net Module                                   &                                \\
                               &                            & Batch Normalization                              &                                \\
                               &                            & MaxPool1D: 5 × 5 pool size, 2 × 2 stride         &                                \\
    \cline{2-4}
                               
                               & Flatten                    & Reshape into 1D vector                           & 0                              \\
    \cline{1-4}
    
\multirow{8}{*}{GFBs}           & \multirow{3}{*}{Block - 1} & BiLSTM: 512 filters                              & \multirow{3}{*}{2109440}       \\
                               &                            & Batch Normalization                              &                                \\
                               &                            & Dropout: 0.1                                     &                                \\
    \cline{2-4}
                               
                               & \multirow{3}{*}{Block - 2} & BiLSTM: 512 filters                              & \multirow{3}{*}{6299648}       \\
                               &                            & Batch Normalization                              &                                \\
                               &                            & Dropout: 0.2                                     &                                \\
    \cline{2-4}
                               
 &
  \multicolumn{1}{l}{\multirow{2}{*}{GlobalAveragePooling1D}} &
  \multirow{2}{*}{Calculates the average value and create a global representation} &
  \multirow{2}{*}{0} \\
                               & \multicolumn{1}{l}{}       &                                                  &                                \\
    \cline{1-4}

Concatenation                  & Block - 1                  & Concatenate both ALFB and GFB                    & 0                              \\
    \cline{1-4}

\multirow{2}{*}{Dense Block} &
  \multirow{2}{*}{Block - 1} &
  Dense: 32 filters, activation = ReLU &
  \multirow{2}{*}{43168} \\
                               &                            & Batch Normalization                              &                                \\
    \cline{1-4}
                               
Output Block                   & Block - 1                  & Activation = Softmax                             & 231                            \\
     \cline{1-4}

\multicolumn{1}{l}{}           & \multicolumn{1}{l}{}       & Total Trainable Parameters                                 & 9137711                        \\ \bottomrule
\end{tabular}%
}
\end{table}

This section focuses on the setup of the proposed model. The specific details of our model are provided in Table \ref{table: structure of propose model}. In order to train the model, we needed to augment the dataset as there were not enough training examples. We used a 60:25:15 ratio to partition the augmented dataset, with 60\% of the data used for training the model, 25\% for validation, and the remaining 15\% for testing the model. We set the batch size to 32 and the learning rate at 0.0001 which is selected by the hyperparameter tuning. The model was trained for 100 epochs in each dataset. For training, the system used an NVIDIA Tesla P100 GPU, 16 GB of RAM, and the TensorFlow backend. The proposed model has a total of 9.14 million trainable parameters. To train the model, we incorporated a loss function by adding categorical cross-entropy. The loss function can be defined as follows:

\begin{equation}
    \label{eq: categorical_crossentropy}
    \begin{split}
        \text{$Loss_{CE}$}&=-\sum_{c=0}^{N} y(i) * log(y(i))
    \end{split}
\end{equation}

The loss function $Loss_{CE}$, as described in Eq. \ref{eq: categorical_crossentropy}, can be defined in terms of the number of classes N, the natural logarithm, and the true probability distribution $y$, along with the predicted probability distribution $\hat{y}$ for each class c. The model adjusts its weights and biases to minimize the categorical cross-entropy loss function. 

\subsection{Hyperparameter Tuning}

For fine-tuning our proposed model, we have chosen Keras Tuner \cite{omalley2019kerastuner} as the hyperparameter tuner. This library is specifically developed for optimizing hyperparameters in Keras. It provides a simple and flexible API for automatically searching the hyperparameters of the deep learning model to enhance its performance. \Cref{tab:alfb_block} represents the hyperparameters used in the ALFBs.


\begin{table}[ht!]
    \scriptsize
    \centering
    \caption{Attention-based Local Feature Blocks (ALFBs)}
    \begin{tabular}{|c|cc|c|}
    \hline
    ALFB & \centering Convolution layer & & 1D Max pooling \\
    \cline{2-3}
    &No of filters&Kernel size&(Pool Size) \\
    \hline
    1st & \multicolumn{1}{l|}{256, 512} & 2, 3, 5 & 2,3,5 \\ 
    \hline
    2nd & \multicolumn{1}{l|}{128, 256, 512} & 2, 3, 5 & 2,3,5 \\ 
    \hline
    3rd & \multicolumn{1}{l|}{64, 128, 256} & 2, 3, 5 & 2,3,5 \\
    \hline
    4th & \multicolumn{1}{l|}{32, 64, 128} & 2, 3, 5 & 2,3,5 \\ 
    \hline
    \end{tabular}

\label{tab:alfb_block}
\end{table}

\begin{table}[h!]
\scriptsize
\centering
\caption{Global Feature Blocks (GFBs)}
\resizebox{0.4\textwidth}{!}{%
    \begin{tabular}{|c|c|c|c|}
    
    \hline
    GFB & BiLSTM (Units) & Dropout \\
    \hline
    1st & 128, 256, 512 & 0.1, 0.2  \\ 
    \hline
    2nd & 128, 256, 512 & 0.1, 0.2  \\ 
    \hline
    \end{tabular}%
}

\label{tab:gfb_block}
\end{table}

\begin{table}[h!]
\scriptsize
\centering
\caption{Dense Block}
\resizebox{0.4\textwidth}{!}{%
\begin{tabular}{|c|c|c|}
    \hline
    Dense Layer & Units \\
    \hline
    1st & 8, 16, 32  \\ 
    \hline
    2nd & Output Labels  \\ 
    \hline
    \end{tabular}%
}

\label{tab:output_block}
\end{table}

We have also tuned two GFBs using the hyperparameters mentioned in Table \ref{tab:gfb_block}. After merging the ALFBs with the GFBs, we passed the sequence through an output block. \Cref{tab:output_block} represents the hyperparameters used in the output block. 

After performing hyperparameter tuning using the Keras Tuner on the utilized datasets, we obtained the best model parameters which improved the overall performance of our model. The results of hyperparameter tuning showed that the ALFBs performed well with 1D convolution layers that had filter sizes of 256, 256, 256, and 32 and kernel sizes of 5, 5, 5, and 3, respectively. The pool size for the 1D max-pooling layers was set to 2, except for the last layer which had a pool size of 5. Based on the results of the hyperparameter tuning, in the GFBs, we found that both the BiLSTM layers performed well with 512 units. For the output blocks, a unit size of 32 was selected for the first dense layer.


We evaluated multiple optimizers, including Adam and RMSprop, and found that Adam performed well for our model. The learning rate of Adam was set to 1e-4 from 1e-2, 1e-3, and 1e-4. We incorporate the resultant output parameter into our model to perform the SER task efficiently.

\section{Result Analysis}
\label{sec: Result Analysis}
In this section, we analyze the achieved results of the proposed model and provide an in-depth comparative analysis with other SOTA existing methods.\\

\subsection{Evaluation Metrics}
In order to measure the performance of our model we have chosen precision, recall, F1 score, and accuracy.

\textbf{Precision}: Precision is a performance measure that evaluates the effectiveness of a classification model \cite{wise1997values}. It quantifies the ratio of true positive predictions to the total number of positive predictions made by the model.

In the form of an equation, precision can be written as:
\begin{equation}
    \begin{split}
        \text{Precision}&=\frac{TP}{TP + FP}
    \end{split}
\end{equation}

Where TP stands for true positives (cases that were correctly labeled as positive) and FP stands for false positives (incorrectly classified positive instances).\\

\textbf{Recall}: The concept of recall \cite{lecompte1994extending} can be represented mathematically as follows:
\begin{equation}
    \begin{split}
        \text{Recall}&=\frac{TP}{TP + FN}
    \end{split}
\end{equation}

Where TP stands for true positives (cases that were correctly labeled as positive), and false negatives, are indicated by FN (incorrectly classified negative instances).\\

\textbf{F1-score}: The F1-score \cite{lipton2014thresholding} is a single metric that provides a balanced measure of the effectiveness of a classification model by combining both precisions and recalls into a single value. The F1 score can be stated as an equation, which looks like this:
\begin{equation}
    \begin{split}
        \text{F1-score}&=\frac{2 * precision * recall}{precision + recall}
    \end{split}
\end{equation}
Where precision and recall are the respective values obtained from the confusion matrix of the classification model.\\

\textbf{K-fold cross-validation}: K-fold cross-validation \cite{anguita2012k} is commonly used to evaluate ML models. It divides the data into k equal-sized segments, trains the model on k-1, and evaluates the remaining subsets. Each subset is tested once. Validation accuracy is used to monitor the performance of the model during training and to modify its hyperparameters. For this study, we have performed a 5-fold cross-validation to evaluate the model's overall performance. Mean accuracy can be expressed as an equation:
\begin{equation}
    \begin{split}
        \text{The mean accuracy}&=\frac{1}{K}\sum_{i=1}^{K}E_i
    \end{split}
\end{equation}

Where, K is the number of folds, and $E_i$ denotes the evaluation metric (such as accuracy, F1-score, etc.) computed for the $i^{th}$ fold. The average of the evaluation metrics across all the folds gives an estimate of the model's performance.

\subsection{Performance Analysis}

\begin{table}[htbp]
\centering
\caption{5-Fold cross-validation analysis using the TESS, RAVDESS, BanglaSER, SUBESCO, and Emo-DB datasets.}
\label{tab:classification_table2}
\resizebox{\textwidth}{!}{%
\begin{tabular}{|c|c|c|c|c|c|c|}
\hline
\textbf{Dataset Name} & \textbf{K-Fold} & \textbf{Accuracy (\%)} & \textbf{Precision (\%)} & \textbf{Recall (\%)} & \textbf{F1-Score (\%)} & \textbf{Mean Accuracy (\%)} \\ \hline
\multirow{5}{*}{TESS}      & 1 & 99.84 & 99.84 & 99.84 & 99.84 & \multirow{5}{*}{99.65} \\ \cline{2-6}
                           & 2 & 99.68 & 99.68 & 99.68 & 99.68 &                        \\ \cline{2-6}
                           & 3 & 99.68 & 99.68 & 99.68 & 99.68 &                        \\ \cline{2-6}
                           & 4 & 99.52 & 99.53 & 99.52 & 99.52 &                        \\ \cline{2-6}
                           & 5 & 99.52 & 99.53 & 99.52 & 99.52 &                        \\ \hline
\multirow{5}{*}{RAVESS}    & 1 & 94.44 & 94.80 & 94.44 & 94.50 & \multirow{5}{*}{94.88} \\ \cline{2-6}
                           & 2 & 96.30 & 96.54 & 96.30 & 96.31 &                        \\ \cline{2-6}
                           & 3 & 95.06 & 95.13 & 95.06 & 95.05 &                        \\ \cline{2-6}
                           & 4 & 94.75 & 94.97 & 94.75 & 94.78 &                        \\ \cline{2-6}
                           & 5 & 93.83 & 94.01 & 93.82 & 93.80 &                        \\ \hline
\multirow{5}{*}{BanglaSER} & 1 & 97.58 & 97.58 & 97.58 & 97.57 & \multirow{5}{*}{98.12} \\ \cline{2-6}
                           & 2 & 97.88 & 97.92 & 97.88 & 97.88 &                        \\ \cline{2-6}
                           & 3 & 98.18 & 98.24 & 98.18 & 98.19 &                        \\ \cline{2-6}
                           & 4 & 98.79 & 98.80 & 98.79 & 98.79 &                        \\ \cline{2-6}
                           & 5 & 98.48 & 98.50 & 98.48 & 98.48 &                        \\ \hline
\multirow{5}{*}{SUBESCO}   & 1 & 97.51 & 97.55 & 97.51 & 97.51 & \multirow{5}{*}{97.94} \\ \cline{2-6}
                           & 2 & 96.97 & 96.99 & 96.98 & 96.98 &                        \\ \cline{2-6}
                           & 3 & 98.75 & 98.51 & 98.48 & 98.49 &                        \\ \cline{2-6}
                           & 4 & 97.42 & 97.4 & 97.42 & 97.42 &                        \\ \cline{2-6}
                           & 5 & 98.04 & 98.06 & 98.04 & 98.04 &                        \\ \hline
\multirow{5}{*}{Emo-DB}    & 1 & 97.52 & 97.70 & 97.52 & 97.55 & \multirow{5}{*}{97.19} \\ \cline{2-6}
                           & 2 & 99.17 & 99.22 & 99.17 & 99.17 &                        \\ \cline{2-6}
                           & 3 & 98.35 & 98.51 & 98.34 & 98.34 &                        \\ \cline{2-6}
                           & 4 & 92.56 & 93.06 & 92.56 & 92.49 &                        \\ \cline{2-6}
                           & 5 & 98.35 & 98.51 & 98.35 & 98.34 &                        \\ \hline
\end{tabular}%
}
\end{table}

\begin{figure}[htpb]
    \centering
    \includegraphics[scale=0.5]{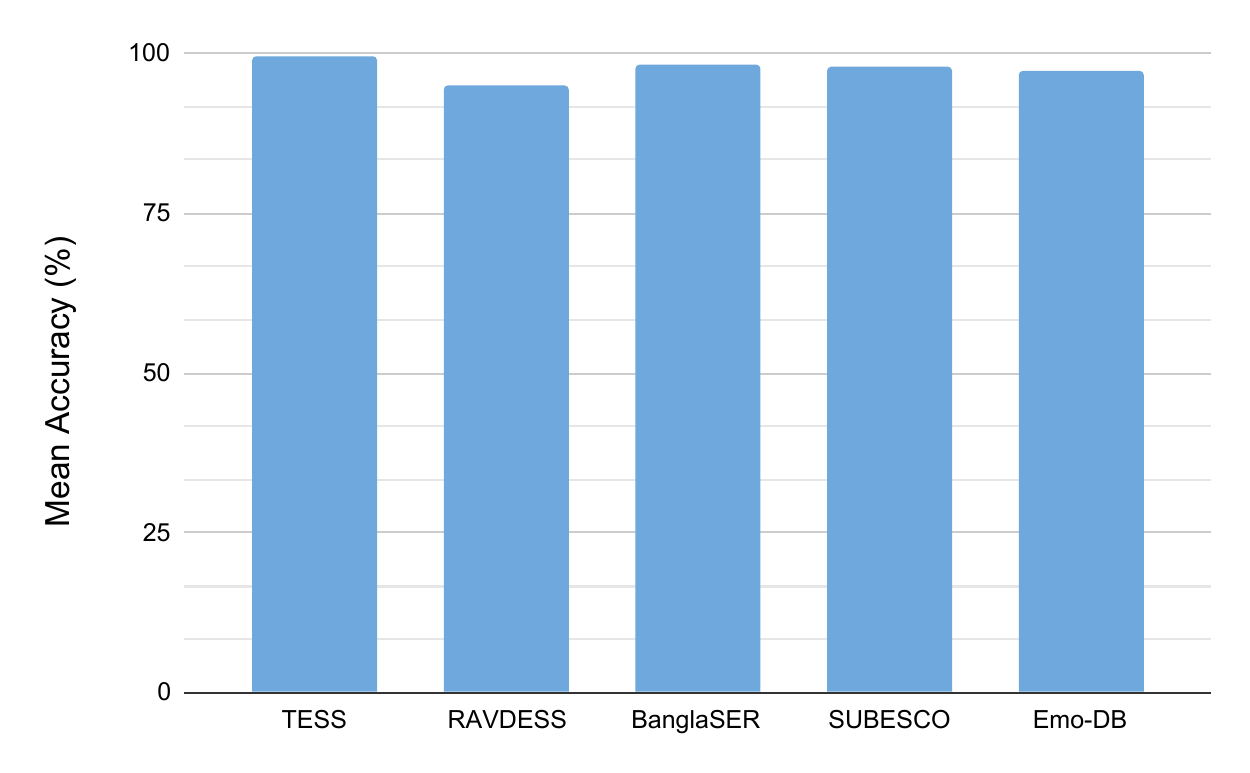}
    \caption{Graphical representation of  the mean accuracy, computed via 5-fold cross-validation, for our proposed model on each of the five datasets.}
    \label{fig:meanaccuracydataset}
\end{figure}

We have utilized five benchmark datasets such as TESS, RAVDESS, BanglaSER, SUBESCO, and Emo-DB. \Cref{tab:classification_table2} represents the five folds cross-validation for each data set and reports validation accuracy, precision, recall, F1-score, and mean accuracy for each dataset. We have taken the weighted average precision, recall, and F1-score as the evaluation metrics for our model. The mean validation accuracy for the TESS dataset is 99.65\%, the mean precision is 99.65\%, the mean recall is 99.65\% and the mean F1-score is 99.65\%. The mean validation accuracy for the RAVDESS dataset is 94.88\%, the mean precision is 95.09\%, the mean recall is 94.87\%, and the mean F1-score is 94.89\%. The mean validation accuracy for the BanglaSER dataset is 98.12\%, mean precision is 98.21\%, mean recall is 98.18\%, and mean F1-score is 98.18\%. The mean validation accuracy for the SUBESCO dataset is 97.94\%, the mean precision is 97.70\%, the mean recall is 97.69\%, and the mean F1-score is 97.69\%. The Emo-DB dataset's mean validation accuracy is 97.19\%, it's mean precision is 97.4\%, mean recall is 97.18\% and it's mean F1-score is 97.18\%. The mean accuracy of the proposed model in all the datasets is presented in Figure \ref{fig:meanaccuracydataset}.

\begin{figure}[htbp]
\centering
    \begin{subfigure}{.45\textwidth}
      \centering
      \captionsetup{justification=centering}
      \includegraphics[scale=0.16]{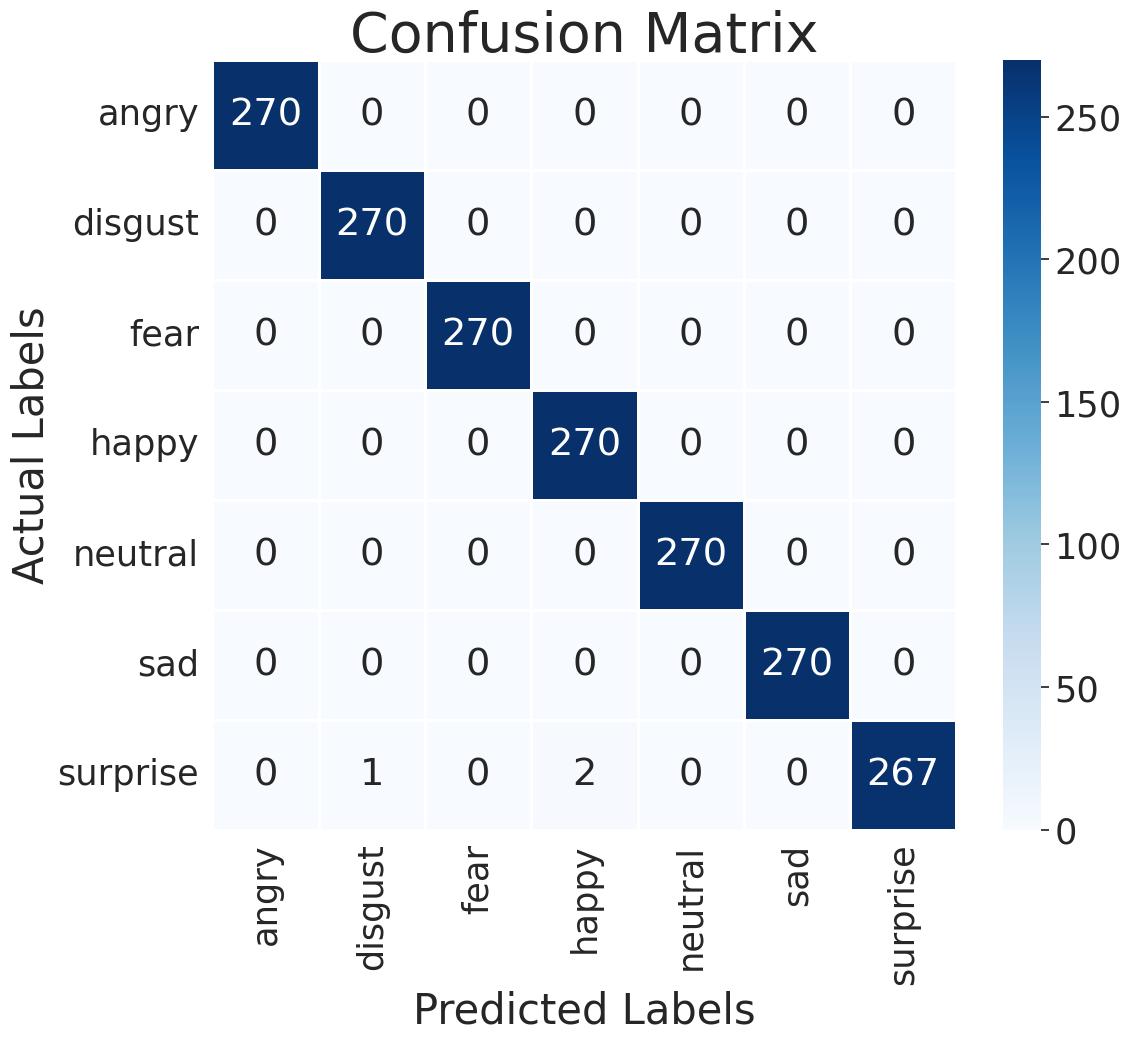}
      \caption{Confusion Matrix of \\ TESS dataset.}
      \label{fig:cm_tess}
    \end{subfigure}
    \begin{subfigure}{.45\textwidth}
      \centering
      \captionsetup{justification=centering}
      \includegraphics[scale=0.16]{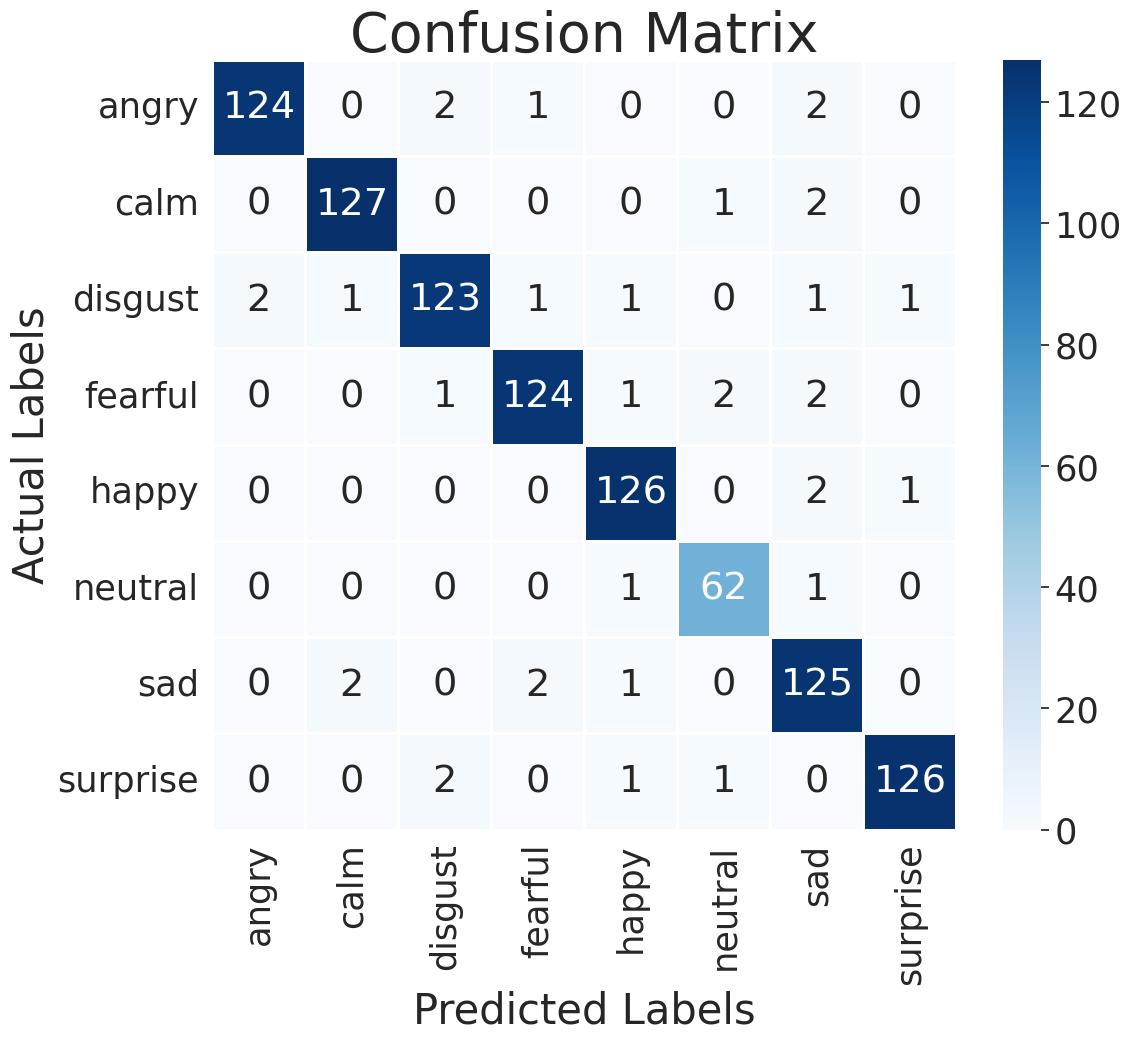}
      \caption{Confusion Matrix of \\ RAVDESS dataset.}
      \label{fig:cm_rav}
    \end{subfigure}
    \par\bigskip
    \begin{subfigure}{.45\textwidth}
       \centering
       \captionsetup{justification=centering}
       \includegraphics[scale=0.18]{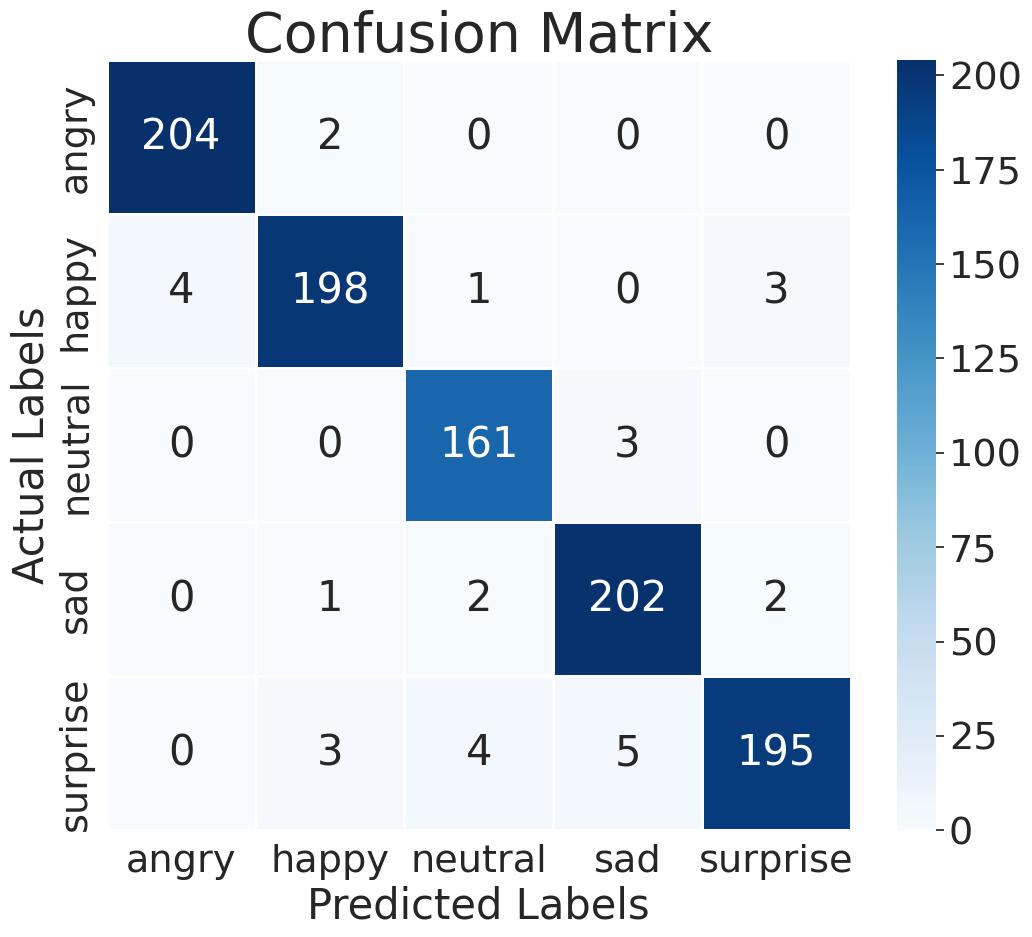}
       \caption{Confusion Matrix of \\ BanglaSER dataset.}
       \label{fig:cm_bangla}
    \end{subfigure}
    \begin{subfigure}{.45\textwidth}
    \centering
    \captionsetup{justification=centering}
       \includegraphics[scale=0.17]{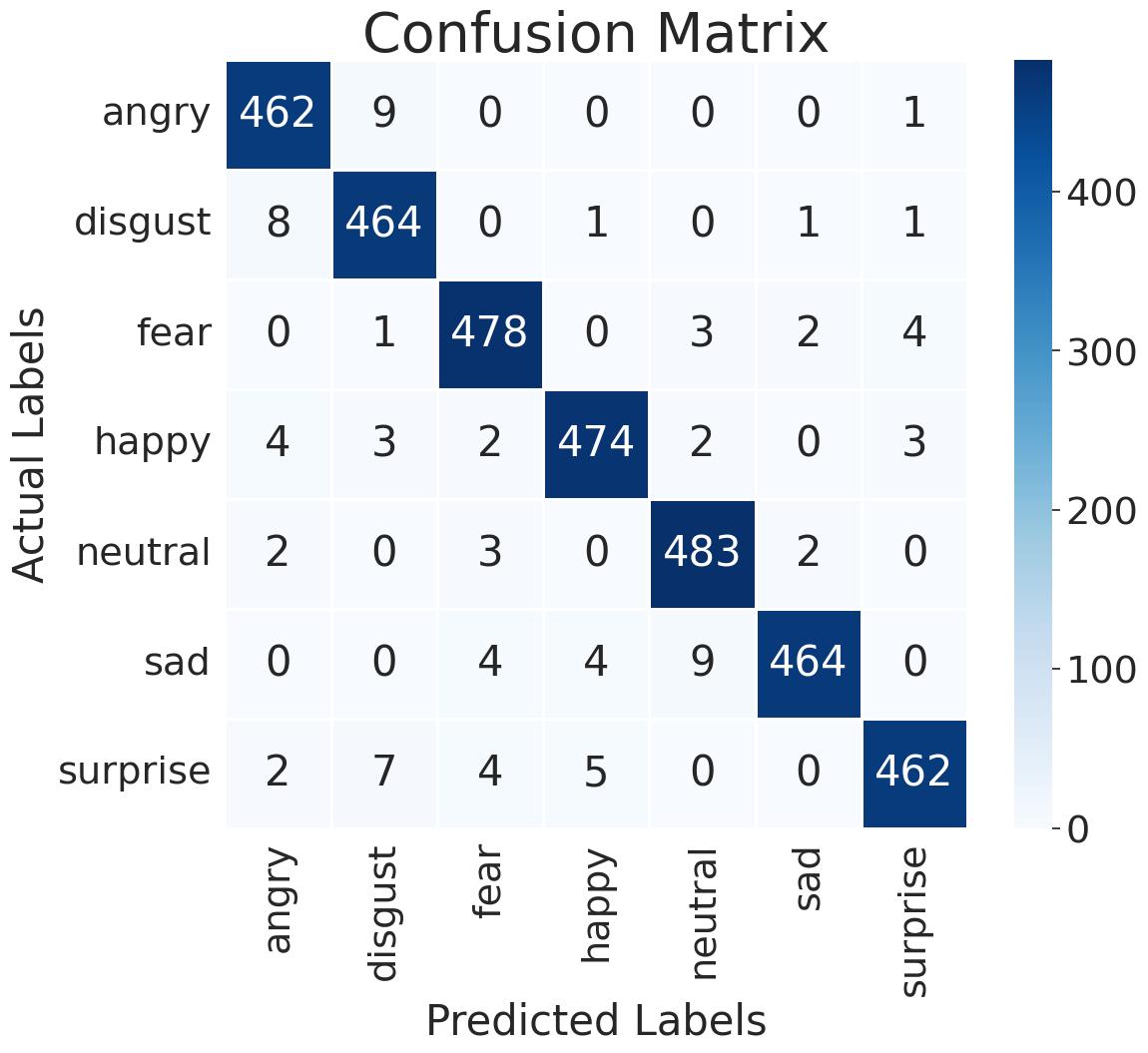}
       \caption{Confusion Matrix of \\ SUBESCO dataset.}
       \label{fig:cm_sub}
    \end{subfigure}
    \par\bigskip
    \begin{subfigure}{.45\textwidth}
    \centering
    \captionsetup{justification=centering}
       \includegraphics[scale=0.17]{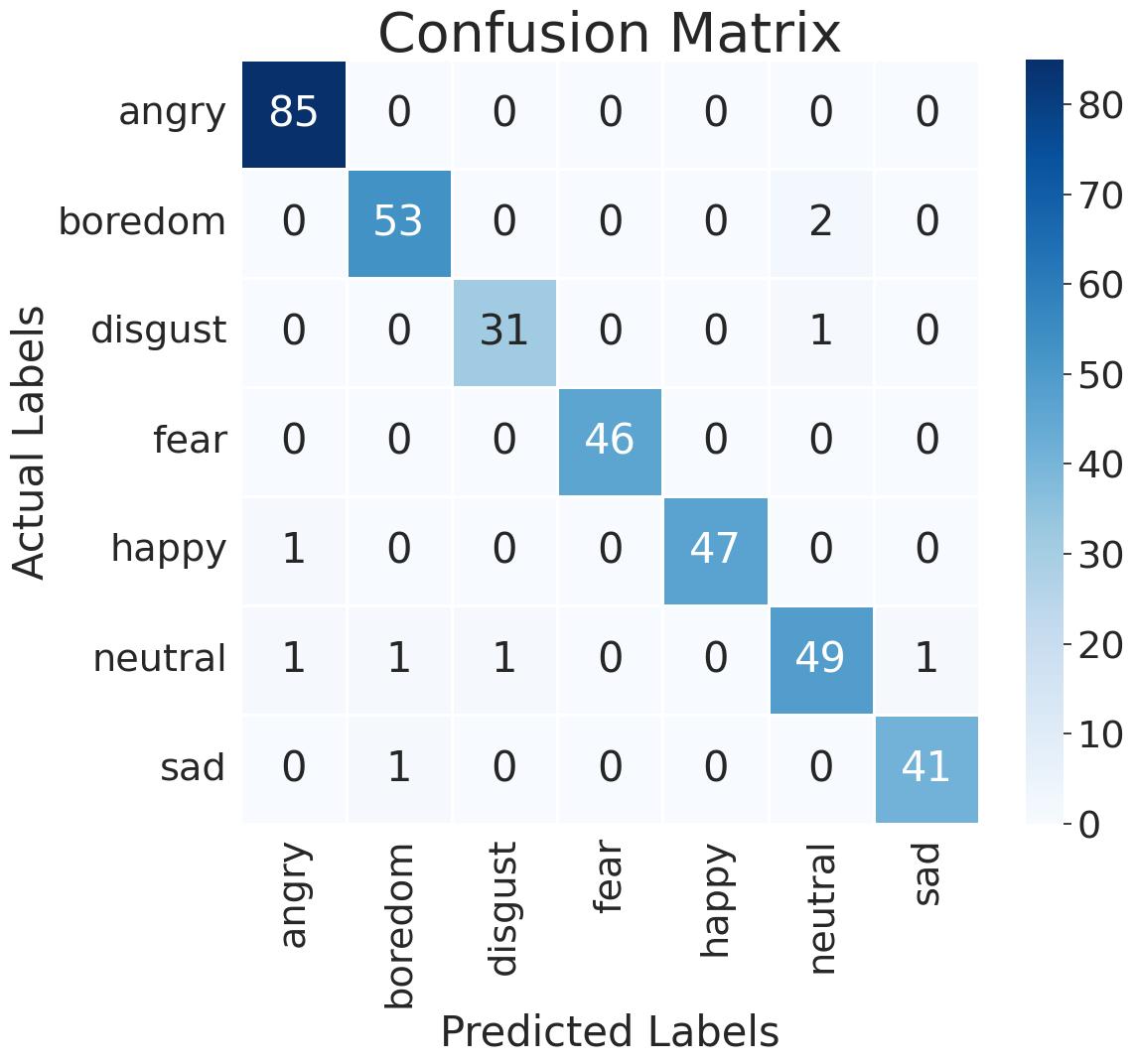}
       \caption{Confusion Matrix of \\ Emo-DB dataset.}
       \label{fig:cm_emodb}
    \end{subfigure}
\caption{Evaluation of the proposed model's performance on the utilized datasets. Here, the confusion matrix has been presented for the
(\subref{fig:cm_tess}) TESS (\subref{fig:cm_rav}) RAVDESS (\subref{fig:cm_bangla}) BanglaSER (\subref{fig:cm_sub}) SUBESCO (\subref{fig:cm_emodb}) datasets of the proposed model.}
\label{fig:confusion_matrix}
\end{figure}

\begin{figure}[htbp]
\centering
    \begin{subfigure}{.45\textwidth}
      \centering
      \captionsetup{justification=centering}
      \includegraphics[scale=0.18]{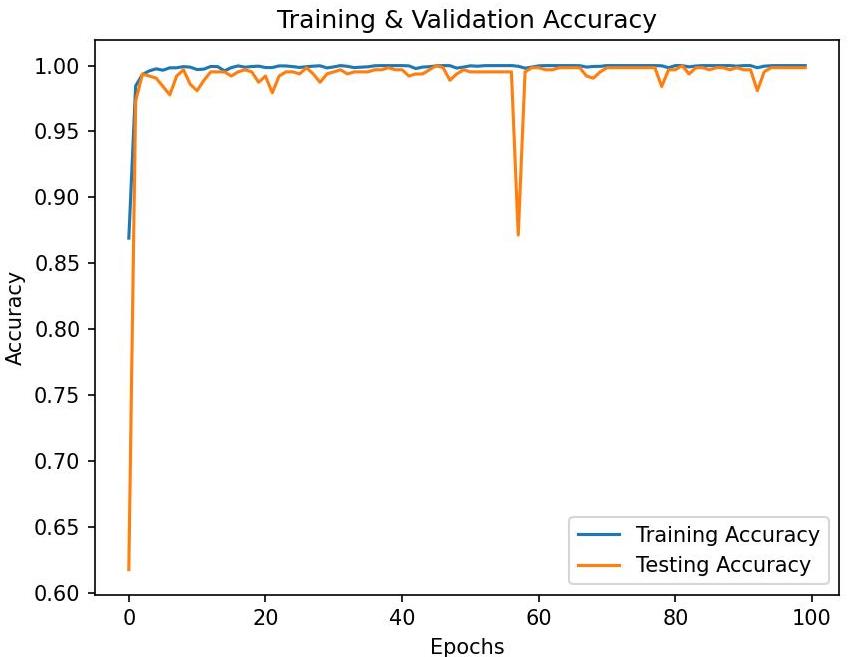}
      \caption{Training vs Validation \\ Accuracy of TESS dataset.}
      \label{fig:tva_tess}
    \end{subfigure}
    \begin{subfigure}{.45\textwidth}
      \centering
      \captionsetup{justification=centering}
      \includegraphics[scale=0.18]{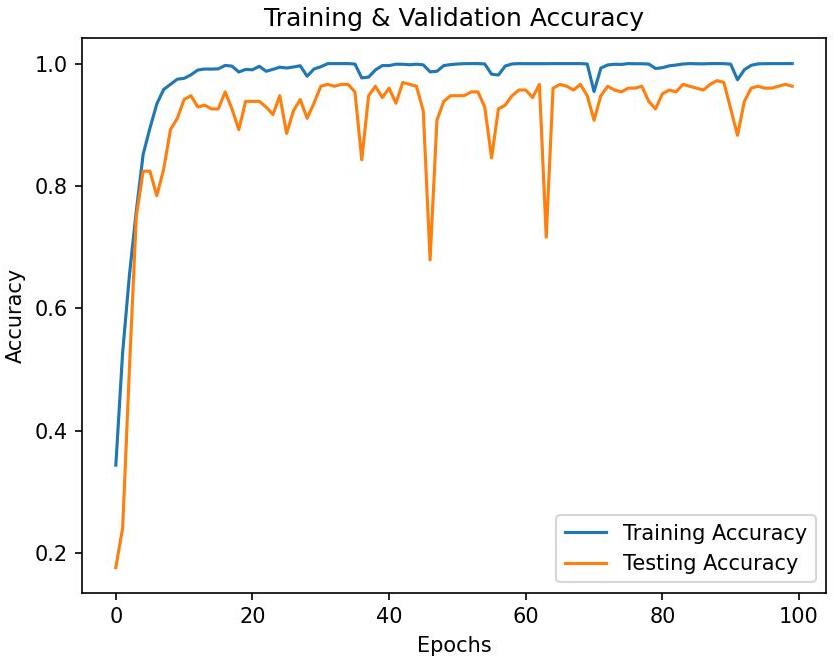}
      \caption{Training vs Validation \\ Accuracy of RAVDESS dataset.}
      \label{fig:tva_rav}
    \end{subfigure}
    \par\bigskip 
    \begin{subfigure}{.45\textwidth}
       \centering
       \captionsetup{justification=centering}
       \includegraphics[scale=0.18]{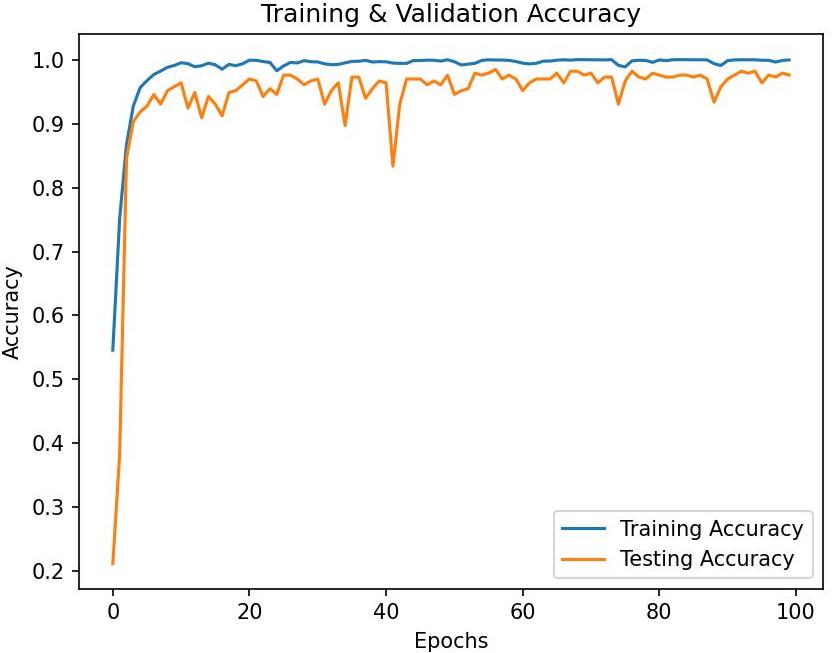}
       \caption{Training vs Validation Accuracy of \\ BanglaSER dataset.}
       \label{fig:tva_bangla}
    \end{subfigure}
    \begin{subfigure}{.45\textwidth}
        \centering
        \captionsetup{justification=centering}
        \includegraphics[scale=0.18]{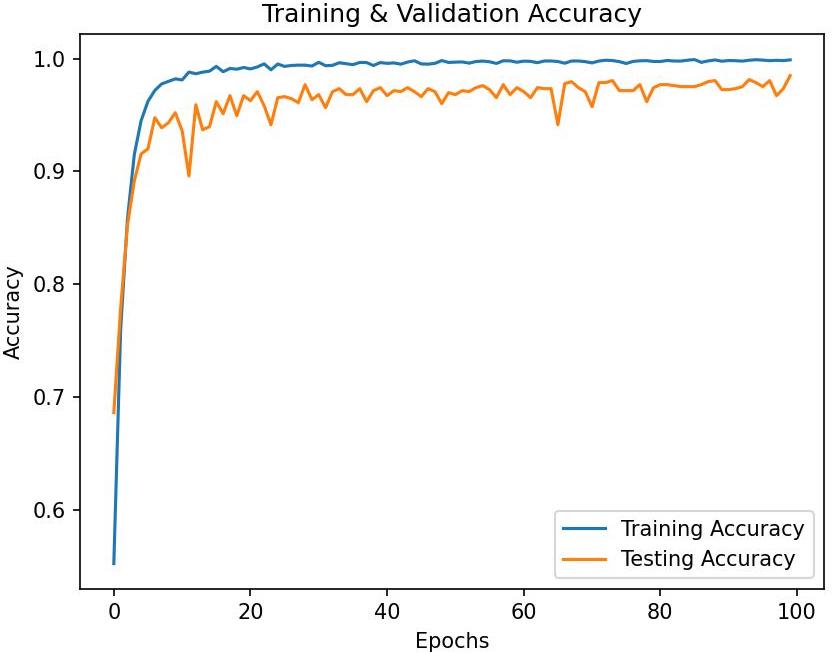}
        \caption{Training vs Validation Accuracy of \\ SUBESCO dataset.}
        \label{fig:tva_sub}
    \end{subfigure}
    \par\bigskip
    \begin{subfigure}{.45\textwidth}
        \centering
        \captionsetup{justification=centering}
       \includegraphics[scale=0.19]{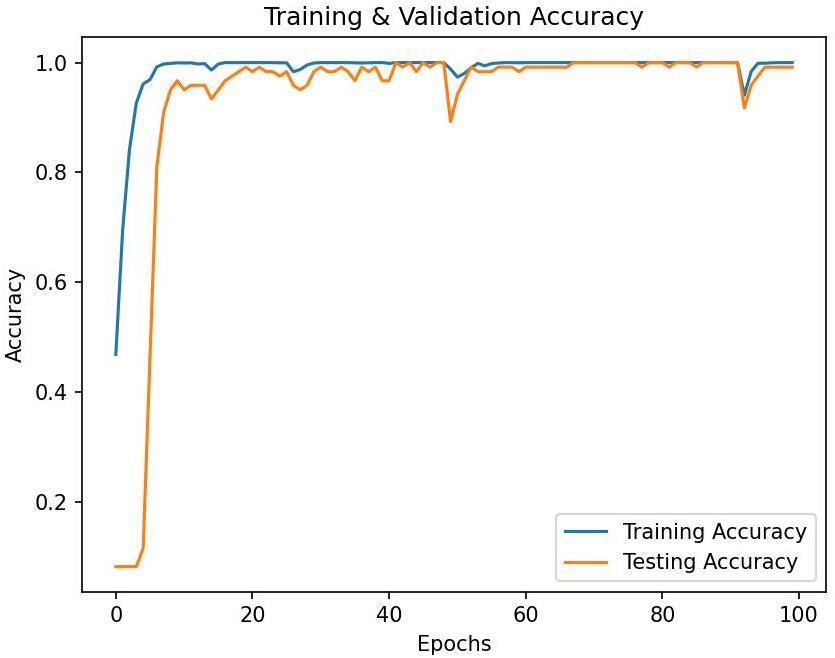}
       \caption{Training vs Validation Accuracy of \\ Emo-DB dataset.}
       \label{fig:tva_emodb}
    \end{subfigure}
\caption{Represent the training and validation loss in our proposed dual-channel multi-lingual speech recognition model on five benchmark datasets, including (\subref{fig:tva_tess}) TESS, (\subref{fig:tva_rav}) RAVDESS, (\subref{fig:tva_bangla}) BanglaSER, (\subref{fig:tva_sub}) SUBESCO and (\subref{fig:tva_emodb})Emo-DB.}
\label{fig:tv_accuracy}
\end{figure}

\begin{figure}[htbp]
\centering
    \begin{subfigure}{.45\textwidth}
      \centering
      \captionsetup{justification=centering}
      \includegraphics[scale=0.17]{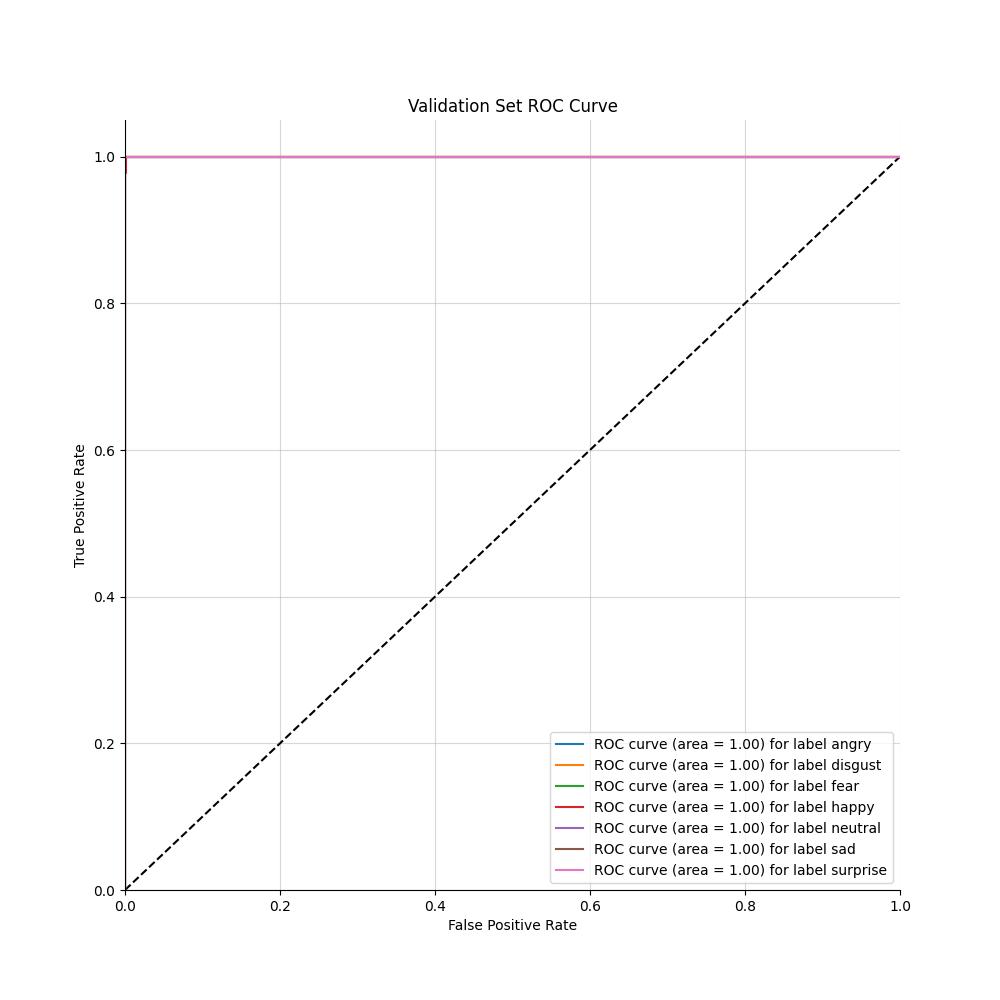}
      \caption{ROC Curve of TESS dataset for \\ the validation dataset.}
      \label{fig:vsroc_tess}
    \end{subfigure}
    \begin{subfigure}{.45\textwidth}
      \centering
      \captionsetup{justification=centering}
      \includegraphics[scale=0.17]{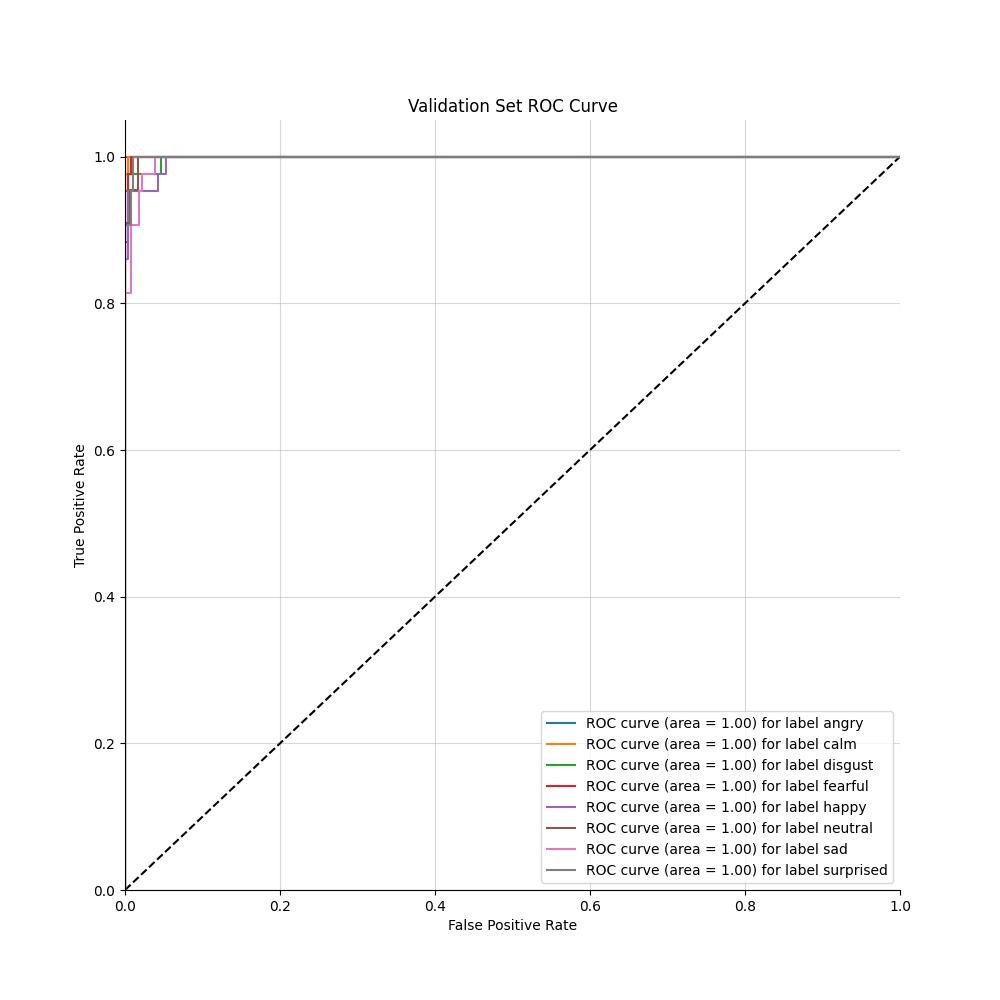}
      \caption{ROC Curve of RAVDESS dataset for \\ the validation dataset.}
      \label{fig:vsroc_rav}
    \end{subfigure}
    \par\bigskip 
    \begin{subfigure}{.45\textwidth}
       \centering
       \captionsetup{justification=centering}
       \includegraphics[scale=0.17]{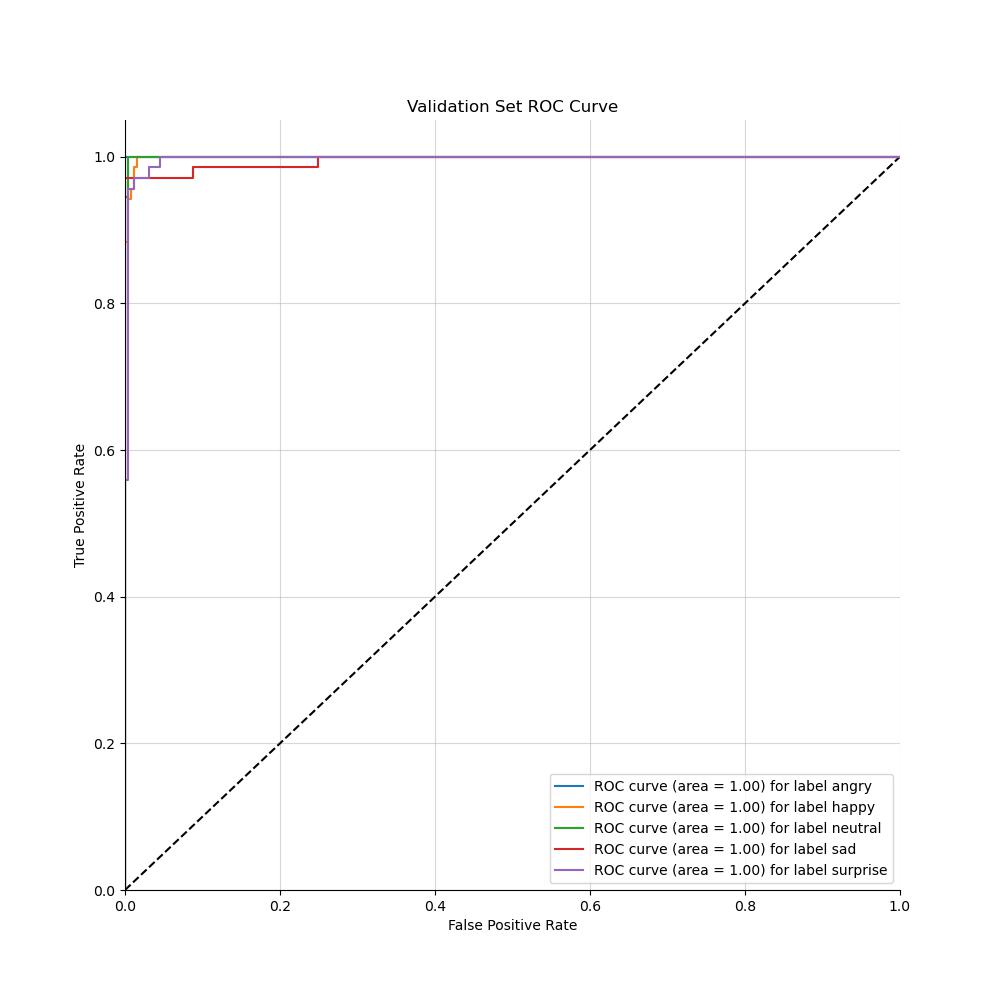}
       \caption{ROC Curve of BanglaSER dataset for \\ the validation dataset.}
       \label{fig:vsroc_bangla}
    \end{subfigure}
    \begin{subfigure}{.45\textwidth}
    \centering
    \captionsetup{justification=centering}
       \includegraphics[scale=0.17]{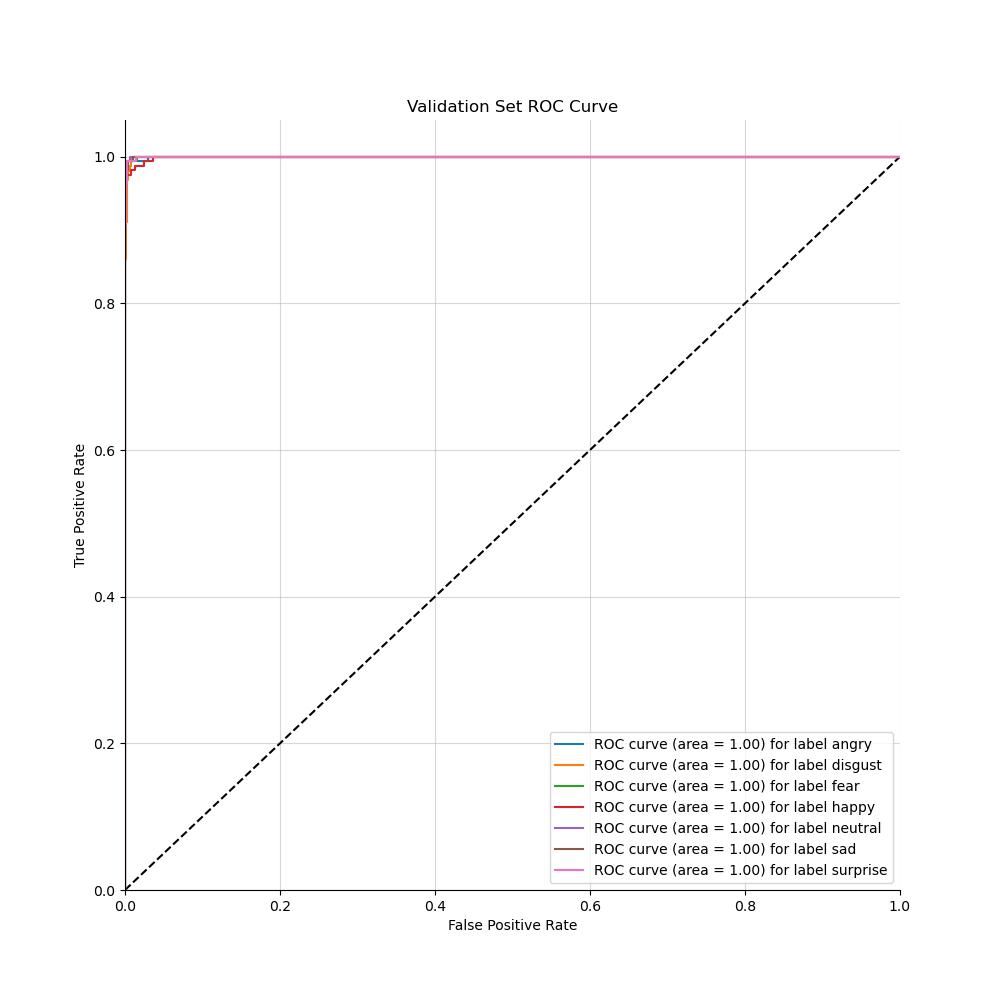}
       \caption{ROC Curve of SUBESCO dataset for \\ the validation dataset.}
       \label{fig:vsroc_sub}
    \end{subfigure}
    \par\bigskip
    \begin{subfigure}{.45\textwidth}
    \centering
    \captionsetup{justification=centering}
       \includegraphics[scale=0.17]{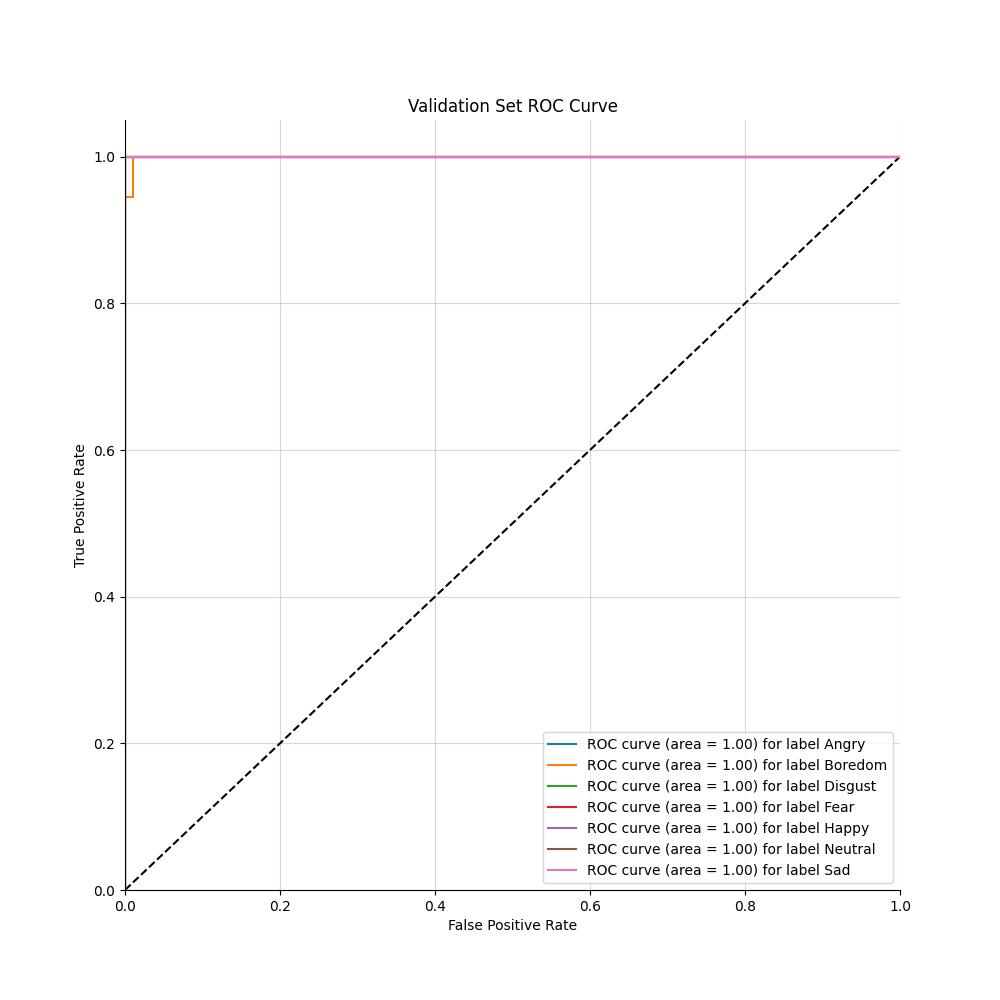}
       \caption{ROC Curve of Emo-DB dataset for \\ the validation dataset.}
       \label{fig:vsroc_emodb}
    \end{subfigure}
\caption{Performance evaluation of the proposed models for (\subref{fig:vsroc_tess}) TESS, (\subref{fig:vsroc_rav}) RAVDESS, (\subref{fig:vsroc_bangla}) BanglaSER, (\subref{fig:vsroc_sub}) SUBESCO and (\subref{fig:vsroc_emodb}) Emo-DB datasets shows the ROC curve of the validation dataset of the proposed model.}
\label{fig:vs_roc}
\end{figure}

The confusion matrix represents the true positive and true negative values along the diagonal. In addition, the off-diagonal entries represent instances of misclassification or confusion between different classes. Figure \ref{fig:confusion_matrix}, represents the confusion matrix of the five benchmark datasets of our proposed model. Our model is comparatively better at classifying different emotion classes. The features of the TESS and Emo-DB datasets are classified well by the proposed model. In most of the datasets, Happy and Sad are perfectly classified by our proposed model.  In the SUBESCO dataset, Disgust shows the highest misclassification rate which is confused with Angry and Surprise. The rest of the datasets also show a good classification result in the confusion matrix. Figure \ref{fig:tv_accuracy} illustrates, the training and validation accuracy achieved in our proposed dual-channel multi-lingual speech recognition model on five benchmark datasets. It has been observed that the training and validation accuracy are closely aligned, indicating the successful fitting of the model. The model has reached stability within 100 epochs, as demonstrated in the graph. Training accuracy measures the accuracy of the model on the training set during the training process, while validation accuracy measures the accuracy of the model on the validation set, which is a subset of the data that was held out from the training process. The alignment of these metrics is a good indicator that our model is not over-fitting or under-fitting the data, and is capable of generalizing well to new and unseen data. Based on this observation, we have run our model for 100 epochs to ensure optimal performance. The Receiver Operating Characteristic (ROC) curve is an important metric used to evaluate the performance of a development model. It can be used for binary or multi-class classification. In binary classification, it is a graphical plot that shows the trade-off between true positive rate and false positive rate, whereas in multi-class classification it shows the diagnostic ability of a classifier for every class. In Figure \ref{fig:vs_roc}, the ROC curve of validation sets from five datasets is illustrated, which was utilized to evaluate the proposed approach. For the TESS dataset, all seven classes are showcasing superior recognition abilities with an Area Under Curve (AUC) score of 1. Similarly, the RAVDESS dataset exhibits significant performance in classifying eight distinct classes. Additionally, great classification performance was showcased by five different emotions in the BanglaSER dataset. Moreover, the SUBESCO dataset, composed of seven different emotions, showcases great distinction ability. Finally, the Emo-DB dataset with seven different types of emotions is displaying impressive classification performance. \\\\\\\\

\subsection{Comparison with State-of-the-art (SOTA) Approaches}

\begin{table}[htbp]
\centering
\caption{Performance analysis of the proposed work and compare it with the latest literature on the TESS dataset.}
\label{table2}
\resizebox{\textwidth}{!}{%
\begin{tabular}{|l|l|l|c|}
\hline
\textbf{Reference}                                                       & \textbf{Methodology}  & \textbf{Features}                             & \textbf{Accuracy} \\ \hline

Dolka et al. \cite{dolka2021speech}             & ANN                          & MFCC

& 	99.52\%                   \\ \hline

Jothimani et al. \cite{jothimani2022mff}               & CNN+LSTM          & MFCC, RMS, ZCR                                   & 99.6 \%           \\

\hline
Uddin et al. \cite{uddin2023efficacy} &
  CNN+LSTM & MFCC, Energy, Fundamental frequency
  &
  97.5\% \\ \hline

Barkur et al. \cite{barkur2022ensemblewave}                     & DNN       & Attention Wavenet                                         & 99\%              \\ \hline
                                                                         &                       & \cellcolor[HTML]{FFFFFF}ZCR, RMS, Chromagram, &                   \\ \cline{3-3}
\multirow{-2}{*}{Ahmed et al. \cite{ahmed2023ensemble}} & \multirow{-2}{*}{1D-CNN-LSTM-GRU} & Log-mel Spectrogram and MFCC & \multirow{-2}{*}{99.46\%} \\ \hline
This work                                                                & 1D CNN+ECA-Net+BiLSTM & RMS, ZCR, MFCC,Mel-spectrogram                & 99.65\%           \\ \hline
\end{tabular}%
}
\end{table}

\begin{table}[htbp]
\centering
\caption{Performance analysis of the proposed work and compare it with the latest literature available on the RAVDESS dataset.}
\label{table3}
\resizebox{\textwidth}{!}{%
\begin{tabular}{|l|l|l|c|}
\hline
\textbf{Reference} &
  \textbf{Methodology} &
  \textbf{Features} &
  \textbf{Accuracy} \\ \hline
\multirow{2}{*}{Sultana et al. \cite{sultana2021bangla}} &
  \multirow{2}{*}{Deep CNN and BiLSTM} &
  MFCCs, pitch, energy &
  \multirow{2}{*}{82.7\%} \\ \cline{3-3}
 &
   &
  and spectral centroid &

\\ \hline
Dair et al. \cite{dair2021linguistic} &
  1D CNN &
  MFCC, CENS, ZCR &
  70.00\% \\ \hline
   Dolka et al. \cite{dolka2021speech} & ANN & MFCC & 88.72\%
        
\\ \hline

Jothimani et al. \cite{jothimani2022mff}               & CNN+LSTM          & MFCC, RMS, ZCR                                   & 92.6 \%     

  \\ \hline
  
\multirow{3}{*}{Kishor Bhangale et al. \cite{bhangale2023speech}} &
  \multirow{3}{*}{1D DCNN} &
  MFCC, LPCC, spectral kurtosis &
  \multirow{3}{*}{94.18\%} \\ \cline{3-3}
 &
   &
  spectrum centroid, spectral roll-off, &
   \\ \cline{3-3}
 &
   &
  wavelet packet transform, ZCR &
   \\ \hline
Alluhaidan et al. \cite{alluhaidan2023speech} &
  CNN &
  MFCC, MFCCT &
  92\% \\ \hline

\multirow{2}{*}{Ahmed et al. \cite{ahmed2023ensemble}} &
  \multirow{2}{*}{1D-CNN-LSTM-GRU} &
  ZCR, RMS, Chromagram &
  \multirow{2}{*}{95.62\%} \\ \cline{3-3}
 &
   &
  Log-mel Spectrogram and MFCC &
   \\ \hline
\multirow{2}{*}{This work} &
  \multirow{2}{*}{1D CNN+ECA-Net+BiLSTM} &
  RMS, ZCR, MFCC &
  \multirow{2}{*}{94.88\%} \\ \cline{3-3}
 &
   &
  Mel-spectrogram &
   \\ \hline
\end{tabular}%
}
\end{table}


\begin{table}[htbp]
\centering
\caption{Performance analysis of the proposed work and compare it with the latest literature available on the BanglaSER dataset.}
\label{table4}
\resizebox{\textwidth}{!}{%
\begin{tabular}{|l|l|l|c|}
\hline
\textbf{Reference}                                    & \textbf{Methodology}  & \textbf{Features}              & \multicolumn{1}{l|}{\textbf{Accuracy}} \\ \hline

Chakraborty et al. \cite{chakraborty2022phase} & CNN                   & Phase-based Cepstral            & 79\%              \\ \hline

This work                                             & 1D CNN+ECA-Net+BiLSTM & RMS, ZCR, MFCC, Mel-spectrogram & 98.12\%                                  \\ \hline
\end{tabular}%
}
\end{table}



\begin{table}[htbp]
\centering
\caption{Performance analysis of the proposed work and compare it with the latest literature available on the SUBESCO dataset.}
\label{table5}
\resizebox{\textwidth}{!}{%
\begin{tabular}{|l|l|l|c|}
\hline
\textbf{Reference}                                              & \textbf{Methodology}  & \textbf{Features}               & \textbf{Accuracy} \\ \hline
Sultana et al. \cite{sultana2021bangla} & Deep CNN and BiLSTM & MFCCs, pitch, energy, and  spectral centroid & 86.9\% \\ \hline
Chakraborty et al. \cite{chakraborty2022phase} & CNN                   & Phase-based Cepstral            & 75\%              \\ \hline
This work                                                       & 1D CNN+ECA-Net+BiLSTM & RMS, ZCR, MFCC, Mel-spectrogram & 97.94\%           \\ \hline
\end{tabular}%
}
\end{table}


\begin{table}[h]
\centering
\caption{Performance analysis of the proposed work and compare it with the latest literature available on the Emo-DB dataset.}
\label{table6}
\resizebox{\textwidth}{!}{%
\begin{tabular}{|l|l|l|c|}
\hline
\textbf{Reference}                                                                & \textbf{Methodology}                & \textbf{Features}                                                & \textbf{Accuracy} \\ \hline

Thakur et al. \cite{thakur2022language} & SVM & Discrete Wavelet
Spectral, Prosodic & 90.02\%
 \\ \hline

Bhattacharya et al. \cite{bhattacharya2022emotion} & CNN & MFCC, chroma, Tonnetz, Contrast &
96.22\%
\\ \hline 

Alluhaidan et al. \cite{alluhaidan2023speech}           & CNN                        & MFCC, MFCCT                                             & 97\%     \\ \hline

\multirow{2}{*}{Al-onazi et al. \cite{al2022transformer} } &
  \multirow{2}{*}{Transformer} &
  Chromagram, Spectral Contrast &
  \multirow{2}{*}{93.40\%} \\ \cline{3-3}
 &
   &
  MFCC, Mel spectrogram, Tonnetz &
   \\ \hline

\multirow{2}{*}{Ahmed et al. \cite{ahmed2023ensemble}} &
  \multirow{2}{*}{1D-CNN-LSTM-GRU} &
  ZCR, RMS, Chromagram, &
  \multirow{2}{*}{95.42\%} \\ \cline{3-3}
                                                                         &                            & Log-mel Spectrogram and MFCC                            &          \\ \hline
This work                                                                & 1D CNN+ECA-Net+BiLSTM      & RMS, ZCR, MFCC, Mel-spectrogram                         & 97.19\%  \\ \hline
\end{tabular}%
}
\end{table}


\Cref{table2,table3,table4,table5,table6} present the results of a performance evaluation conducted on five different datasets. Our main objective is to compare the effectiveness of our proposed model against previous SOTA approaches. To ensure a fair comparison, we have prioritized the classification performance of the model over computational efficiency, taking into account factors such as the methodology employed, the extracted features, and the accuracy of the approaches. Results presented in Table \ref{table2}, demonstrate that our proposed method outperforms SOTA architectures on the TESS dataset, achieving an impressive classification accuracy of 99.65\%. Our model surpasses all SOTA architectures in this dataset. In Table \ref{table3}, when considering the RAVDESS dataset, the 1D-CNN-LSTM-GRU model \cite{ahmed2023ensemble} achieves the highest accuracy among the other models, reaching 95.62\%. However, our model performs significantly better compared to the existing models in this dataset. Turning to the BanglaSER dataset in Table \ref{table4}, our proposed model achieves significantly improved results compared to all SOTA architectures. For a fair comparison, we did not include several papers \cite{hasan2020emotion} \cite{meng2019speech} \cite{wang2021novel} that classified six or four separate categories whereas we used five. In the SUBESCO dataset in Table \ref{table5}, our proposed model achieved an accuracy of 97.94\%, showcasing a substantial improvement of 11.04\% and 22.94\% compared to the DCNN and BiLSTM \cite{sultana2021bangla} as well as the CNN \cite{chakraborty2022phase} architectures, respectively. Lastly, our proposed model achieves an accuracy of 97.19\% on the Emo-DB dataset, as shown in Table \ref{table6}.

\subsection{Ablation Study}

We have conducted an extensive ablation study in the Emo-DB and RAVDESS datasets to examine the impact of incorporated components such as ECA-Net in the ALFBs, as well as the number of incorporated GFBs. Table \ref{tab:ablation_table}, represents the experimental results from the ablation study. Here, the proposed model shows the mean validation accuracy with the mean precision, recall, and F1 score. The rest of the experiments in the ablation study was run two times. Each result shows the mean value from the experiments. The study demonstrates that, if we remove the ECA-Net and GFBs from the proposed architecture, the performance of the proposed model tends to reduce. In the Emo-DB dataset, the proposed architecture without ECA-Net and GFBs achieved an accuracy of 95.04\%. Removal of the ECA-Net module from ALFBs reduced the mean accuracy by 0.84\%. This observation emphasizes the significance of the ECA-Net module in enhancing the model's performance. Furthermore, when the proposed model was tested without one GFB, the accuracy of the model is 97.52\%. Removal of GFBs altogether from the model's architecture reduced the recognition performance by 0.44\%. 

\begin{table}[htbp]
\scriptsize
\caption{Experimental results from the ablation study.}
\label{tab:ablation_table}
\resizebox{\textwidth}{!}{%
\begin{tabular}{|l|l|l|l|l|l|}
\hline
\textbf{Dataset} & \textbf{Model}                                         & \textbf{Accuracy} \textbf{(\%)} & \textbf{Precision} \textbf{(\%)} & \textbf{Recall} \textbf{(\%)} & \textbf{F1-score} \textbf{(\%)} \\ \hline
\multirow{5}{*}{Emo-DB}  
        & Proposed Architecture w/o ECA-Net \& GFBs & 95.04 & 95.33 & 95.04 & 94.99 \\ \cline{2-6}  
        & Proposed Architecture w/o GFBs & 96.75    & 96.80     & 96.75  & 96.74    \\ \cline{2-6} 
        & Proposed Architecture w/o ECA-Net & 96.35    & 96.36     & 96.35  & 96.35    \\ \cline{2-6}
        & Proposed Architecture w/o One GFB & 97.01    & 97.01     & 97.01  & 97.01    \\ \cline{2-6} 
        & Proposed Architecture & 97.19 & 97.40 & 97.18 & 97.18 \\ 
        \hline
\multirow{5}{*}{RAVDESS} 
        & Proposed Architecture w/o ECA-Net \& GFBs & 91.97 & 92.33 & 91.97 & 91.99  \\ \cline{2-6} 
        & Proposed Architecture w/o GFBs & 93.89    & 93.90     & 93.88  & 93.88    \\ \cline{2-6} 
        & Proposed Architecture w/o ECA-Net & 93.04    & 93.88     & 93.04  & 93.04    \\ \cline{2-6} 
        & Proposed Architecture w/o One GFB & 94.45    & 94.44     & 94.45  & 94.45    \\ \cline{2-6} 
        & Proposed Architecture & 94.88     & 95.09     & 94.87   & 94.89    \\ 
        \hline
\end{tabular}
}
\end{table}

In the RAVDESS dataset, we observed a similar pattern in all the evaluation metrics, as shown in Table \ref{tab:ablation_table}. Removing the ECA-Net as well as GFBs from the proposed architecture reduced the accuracy by 2.91\%. The absence of GFBs and ECA-Net contributed to the loss of accuracy of 0.99\% and 1.84\%, respectively in the RAVDESS dataset. Comparing the performance of the model without ECA-Net and GFBs as well as the model without ECA-Net shows that both of these integrated modules contribute highly to the overall performance.

\section{Conclusion}
In this paper, a dual-channel attention-guided SER architecture is proposed that tries to address the challenge of efficient feature representation in the presence of limited data availability. The proposed model's integration of ECA-Net, 1D CNN, and BiLSTM network performs better in representing the local salient features as well as global contextual features of emotional speech utterances. Integration of ECA-Net allows the network on capturing salient and interrelated features between different channels. On the other hand, BiLSTM is integrated to capture global features from speech data. Data augmentation techniques were employed to mitigate the issue of data scarcity and enhance classification performance. Through comprehensive evaluations of five standard benchmark datasets from multiple languages, our proposed model demonstrated SOTA results compared to other studies. Future prospects involve devising a multi-modal architecture that incorporates audio, video, and textual data while maintaining a lower computation cost.




\bibliographystyle{elsarticle-num} 
\bibliography{cas-refs}

\begin{thebibliography}{10}
\expandafter\ifx\csname url\endcsname\relax
  \def\url#1{\texttt{#1}}\fi
\expandafter\ifx\csname urlprefix\endcsname\relax\def\urlprefix{URL }\fi
\expandafter\ifx\csname href\endcsname\relax
  \def\href#1#2{#2} \def\path#1{#1}\fi

\bibitem{el2011survey}
M.~El~Ayadi, M.~S. Kamel, F.~Karray, Survey on speech emotion recognition: Features, classification schemes, and databases, Pattern recognition 44~(3) (2011) 572--587.

\bibitem{sultana2021bangla}
S.~Sultana, M.~Z. Iqbal, M.~R. Selim, M.~M. Rashid, M.~S. Rahman, Bangla speech emotion recognition and cross-lingual study using deep cnn and blstm networks, IEEE Access 10 (2021) 564--578.

\bibitem{france2000acoustical}
D.~J. France, R.~G. Shiavi, S.~Silverman, M.~Silverman, M.~Wilkes, Acoustical properties of speech as indicators of depression and suicidal risk, IEEE transactions on Biomedical Engineering 47~(7) (2000) 829--837.

\bibitem{n1}
X.~Li, J.~Tao, M.~T. Johnson, J.~Soltis, A.~Savage, K.~M. Leong, J.~D. Newman, Stress and emotion classification using jitter and shimmer features, in: 2007 IEEE International Conference on Acoustics, Speech and Signal Processing-ICASSP'07, Vol.~4, IEEE, 2007, pp. IV--1081.

\bibitem{eyben2010towards}
F.~Eyben, A.~Batliner, B.~Schuller, Towards a standard set of acoustic features for the processing of emotion in speech., in: Proceedings of Meetings on Acoustics 159ASA, Vol.~9, Acoustical Society of America, 2010, p. 060006.

\bibitem{eyben2015geneva}
F.~Eyben, K.~R. Scherer, B.~W. Schuller, J.~Sundberg, E.~Andr{\'e}, C.~Busso, L.~Y. Devillers, J.~Epps, P.~Laukka, S.~S. Narayanan, et~al., The geneva minimalistic acoustic parameter set (gemaps) for voice research and affective computing, IEEE transactions on affective computing 7~(2) (2015) 190--202.

\bibitem{ahmed2023ensemble}
M.~R. Ahmed, S.~Islam, A.~M. Islam, S.~Shatabda, An ensemble 1d-cnn-lstm-gru model with data augmentation for speech emotion recognition, Expert Systems with Applications 218 (2023) 119633.

\bibitem{bhavan2019bagged}
A.~Bhavan, P.~Chauhan, R.~R. Shah, et~al., Bagged support vector machines for emotion recognition from speech, Knowledge-Based Systems 184 (2019) 104886.

\bibitem{8683172}
S.~Mao, D.~Tao, G.~Zhang, P.~C. Ching, T.~Lee, Revisiting hidden markov models for speech emotion recognition, in: ICASSP 2019 - 2019 IEEE International Conference on Acoustics, Speech and Signal Processing (ICASSP), 2019, pp. 6715--6719.
\newblock \href {https://doi.org/10.1109/ICASSP.2019.8683172} {\path{doi:10.1109/ICASSP.2019.8683172}}.

\bibitem{8651287}
I.~Shahin, A.~B. Nassif, S.~Hamsa, Emotion recognition using hybrid gaussian mixture model and deep neural network, IEEE Access 7 (2019) 26777--26787.
\newblock \href {https://doi.org/10.1109/ACCESS.2019.2901352} {\path{doi:10.1109/ACCESS.2019.2901352}}.

\bibitem{venkata2021speech}
M.~Venkata~Subbarao, S.~K. Terlapu, N.~Geethika, K.~D. Harika, Speech emotion recognition using k-nearest neighbor classifiers, in: Recent Advances in Artificial Intelligence and Data Engineering: Select Proceedings of AIDE 2020, Springer, 2021, pp. 123--131.

\bibitem{sun2019speech}
L.~Sun, B.~Zou, S.~Fu, J.~Chen, F.~Wang, Speech emotion recognition based on dnn-decision tree svm model, Speech Communication 115 (2019) 29--37.

\bibitem{8462677}
P.~Tzirakis, J.~Zhang, B.~W. Schuller, End-to-end speech emotion recognition using deep neural networks, in: 2018 IEEE International Conference on Acoustics, Speech and Signal Processing (ICASSP), 2018, pp. 5089--5093.
\newblock \href {https://doi.org/10.1109/ICASSP.2018.8462677} {\path{doi:10.1109/ICASSP.2018.8462677}}.

\bibitem{xu2022multi}
X.~Xu, D.~Li, Y.~Zhou, Z.~Wang, Multi-type features separating fusion learning for speech emotion recognition, Applied Soft Computing 130 (2022) 109648.

\bibitem{chen2023learning}
Z.~Chen, J.~Li, H.~Liu, X.~Wang, H.~Wang, Q.~Zheng, Learning multi-scale features for speech emotion recognition with connection attention mechanism, Expert Systems with Applications 214 (2023) 118943.

\bibitem{lotfian2018predicting}
R.~Lotfian, C.~Busso, Predicting categorical emotions by jointly learning primary and secondary emotions through multitask learning, Interspeech 2018 (2018).

\bibitem{lecun2015deep}
Y.~LeCun, Y.~Bengio, G.~Hinton, Deep learning, nature 521~(7553) (2015) 436--444.

\bibitem{chen2015convolutional}
Y.~Chen, Convolutional neural network for sentence classification, Master's thesis, University of Waterloo (2015).

\bibitem{vrysis2020experimenting}
L.~Vrysis, I.~Thoidis, C.~Dimoulas, G.~Papanikolaou, Experimenting with 1d cnn architectures for generic audio classification, in: Audio Engineering Society Convention 148, Audio Engineering Society, 2020.

\bibitem{graves2006connectionist}
A.~Graves, S.~Fern{\'a}ndez, F.~Gomez, J.~Schmidhuber, Connectionist temporal classification: labelling unsegmented sequence data with recurrent neural networks, in: Proceedings of the 23rd international conference on Machine learning, 2006, pp. 369--376.

\bibitem{cho2014learning}
K.~Cho, B.~Van~Merri{\"e}nboer, C.~Gulcehre, D.~Bahdanau, F.~Bougares, H.~Schwenk, Y.~Bengio, Learning phrase representations using rnn encoder-decoder for statistical machine translation, arXiv preprint arXiv:1406.1078 (2014).

\bibitem{hochreiter1997long}
S.~Hochreiter, J.~Schmidhuber, Long short-term memory, Neural computation 9~(8) (1997) 1735--1780.

\bibitem{schuster1997bidirectional}
M.~Schuster, K.~K. Paliwal, Bidirectional recurrent neural networks, IEEE transactions on Signal Processing 45~(11) (1997) 2673--2681.

\bibitem{chen2021stock}
Y.~Chen, R.~Fang, T.~Liang, Z.~Sha, S.~Li, Y.~Yi, W.~Zhou, H.~Song, Stock price forecast based on cnn-bilstm-eca model, Scientific Programming 2021 (2021) 1--20.

\bibitem{wang2020eca}
Q.~Wang, B.~Wu, P.~Zhu, P.~Li, W.~Zuo, Q.~Hu, Eca-net: Efficient channel attention for deep convolutional neural networks, in: Proceedings of the IEEE/CVF conference on computer vision and pattern recognition, 2020, pp. 11534--11542.

\bibitem{dietterich2002ensemble}
T.~G. Dietterich, et~al., Ensemble learning, The handbook of brain theory and neural networks 2~(1) (2002) 110--125.

\bibitem{dupuis2015aging}
K.~Dupuis, M.~K. Pichora-Fuller, Aging affects identification of vocal emotions in semantically neutral sentences, Journal of Speech, Language, and Hearing Research 58~(3) (2015) 1061--1076.

\bibitem{livingstone2018ryerson}
S.~R. Livingstone, F.~A. Russo, The ryerson audio-visual database of emotional speech and song (ravdess): A dynamic, multimodal set of facial and vocal expressions in north american english, PloS one 13~(5) (2018) e0196391.

\bibitem{das2022banglaser}
R.~K. Das, N.~Islam, M.~R. Ahmed, S.~Islam, S.~Shatabda, A.~M. Islam, Banglaser: A speech emotion recognition dataset for the bangla language, Data in Brief 42 (2022) 108091.

\bibitem{sultana2021sust}
S.~Sultana, M.~S. Rahman, M.~R. Selim, M.~Z. Iqbal, Sust bangla emotional speech corpus (subesco): An audio-only emotional speech corpus for bangla, Plos one 16~(4) (2021) e0250173.

\bibitem{burkhardt2005database}
F.~Burkhardt, A.~Paeschke, M.~Rolfes, W.~F. Sendlmeier, B.~Weiss, et~al., A database of german emotional speech., in: Interspeech, Vol.~5, 2005, pp. 1517--1520.

\bibitem{A2}
S.~Kwon, et~al., Mlt-dnet: Speech emotion recognition using 1d dilated cnn based on multi-learning trick approach, Expert Systems with Applications 167 (2021) 114177.

\bibitem{alluhaidan2023speech}
A.~S. Alluhaidan, O.~Saidani, R.~Jahangir, M.~A. Nauman, O.~S. Neffati, Speech emotion recognition through hybrid features and convolutional neural network, Applied Sciences 13~(8) (2023) 4750.

\bibitem{patnaik2023speech}
S.~Patnaik, Speech emotion recognition by using complex mfcc and deep sequential model, Multimedia Tools and Applications 82~(8) (2023) 11897--11922.

\bibitem{zhong2023speech}
Z.~Zhong, Speech emotion recognition based on svm and cnn using mfcc feature extraction, in: International Conference on Statistics, Data Science, and Computational Intelligence (CSDSCI 2022), Vol. 12510, SPIE, 2023, pp. 445--452.

\bibitem{A13}
A.~Aggarwal, A.~Srivastava, A.~Agarwal, N.~Chahal, D.~Singh, A.~A. Alnuaim, A.~Alhadlaq, H.-N. Lee, Two-way feature extraction for speech emotion recognition using deep learning, Sensors 22~(6) (2022) 2378.

\bibitem{simonyan2014very}
K.~Simonyan, A.~Zisserman, Very deep convolutional networks for large-scale image recognition, arXiv preprint arXiv:1409.1556 (2014).

\bibitem{hu2018squeeze}
J.~Hu, L.~Shen, G.~Sun, Squeeze-and-excitation networks, in: Proceedings of the IEEE conference on computer vision and pattern recognition, 2018, pp. 7132--7141.

\bibitem{woo2018cbam}
S.~Woo, J.~Park, J.-Y. Lee, I.~S. Kweon, Cbam: Convolutional block attention module, in: Proceedings of the European conference on computer vision (ECCV), 2018, pp. 3--19.

\bibitem{A3}
H.~Zou, Y.~Si, C.~Chen, D.~Rajan, E.~S. Chng, Speech emotion recognition with co-attention based multi-level acoustic information, in: ICASSP 2022-2022 IEEE International Conference on Acoustics, Speech and Signal Processing (ICASSP), IEEE, 2022, pp. 7367--7371.

\bibitem{zhao2021combining}
Z.~Zhao, Q.~Li, Z.~Zhang, N.~Cummins, H.~Wang, J.~Tao, B.~W. Schuller, Combining a parallel 2d cnn with a self-attention dilated residual network for ctc-based discrete speech emotion recognition, Neural Networks 141 (2021) 52--60.

\bibitem{mustaqeem2021speech}
M.~Mustaqeem, S.~Kwon, Speech emotion recognition based on deep networks: A review, in: Proceedings of the Korea Information Processing Society Conference, Korea Information Processing Society, 2021, pp. 331--334.

\bibitem{A9}
L.~Sun, B.~Liu, J.~Tao, Z.~Lian, Multimodal cross-and self-attention network for speech emotion recognition, in: ICASSP 2021-2021 IEEE International Conference on Acoustics, Speech and Signal Processing (ICASSP), IEEE, 2021, pp. 4275--4279.

\bibitem{zhang2021speech}
Z.~Zhang, Speech feature selection and emotion recognition based on weighted binary cuckoo search, Alexandria Engineering Journal 60~(1) (2021) 1499--1507.

\bibitem{falahzadeh2023deep}
M.~R. Falahzadeh, F.~Farokhi, A.~Harimi, R.~Sabbaghi-Nadooshan, Deep convolutional neural network and gray wolf optimization algorithm for speech emotion recognition, Circuits, Systems, and Signal Processing 42~(1) (2023) 449--492.

\bibitem{chalapathi2022ensemble}
M.~Chalapathi, M.~R. Kumar, N.~Sharma, S.~Shitharth, Ensemble learning by high-dimensional acoustic features for emotion recognition from speech audio signal, Security and Communication Networks 2022 (2022).

\bibitem{A11}
K.~Zvarevashe, O.~O. Olugbara, Recognition of cross-language acoustic emotional valence using stacked ensemble learning, Algorithms 13~(10) (2020) 246.

\bibitem{awgn}
C.~Shannon, Communication in the presence of noise, Proceedings of the IRE 37~(1) (1949) 10--21.
\newblock \href {https://doi.org/10.1109/JRPROC.1949.232969} {\path{doi:10.1109/JRPROC.1949.232969}}.

\bibitem{dolka2021speech}
H.~Dolka, A.~X. VM, S.~Juliet, Speech emotion recognition using ann on mfcc features, in: 2021 3rd international conference on signal processing and communication (ICPSC), IEEE, 2021, pp. 431--435.

\bibitem{liu2018speech}
Z.-T. Liu, Q.~Xie, M.~Wu, W.-H. Cao, Y.~Mei, J.-W. Mao, Speech emotion recognition based on an improved brain emotion learning model, Neurocomputing 309 (2018) 145--156.

\bibitem{leitner2019audio}
M.~Leitner, J.~Thornton, Audio recognition using mel spectrograms and convolution neural networks, arXiv preprint arXiv:1905.00078 (2019).

\bibitem{lu2020speech}
Q.~Lu, Y.~Li, Z.~Qin, X.~Liu, Y.~Xie, Speech recognition using efficientnet, in: Proceedings of the 2020 5th International Conference on Multimedia Systems and Signal Processing, 2020, pp. 64--68.

\bibitem{koduru2020feature}
A.~Koduru, H.~B. Valiveti, A.~K. Budati, Feature extraction algorithms to improve the speech emotion recognition rate, International Journal of Speech Technology 23~(1) (2020) 45--55.

\bibitem{er2020novel}
M.~B. Er, A novel approach for classification of speech emotions based on deep and acoustic features, IEEE Access 8 (2020) 221640--221653.

\bibitem{bhangale2023speech}
K.~Bhangale, M.~Kothandaraman, Speech emotion recognition based on multiple acoustic features and deep convolutional neural network, Electronics 12~(4) (2023) 839.

\bibitem{omalley2019kerastuner}
T.~O'Malley, E.~Bursztein, J.~Long, F.~Chollet, H.~Jin, L.~Invernizzi, et~al., Kerastuner, \url{https://github.com/keras-team/keras-tuner} (2019).

\bibitem{wise1997values}
M.~N. Wise, The values of precision, Princeton University Press, 1997.

\bibitem{lecompte1994extending}
D.~C. LeCompte, Extending the irrelevant speech effect beyond serial recall., Journal of Experimental Psychology: Learning, Memory, and Cognition 20~(6) (1994) 1396.

\bibitem{lipton2014thresholding}
Z.~C. Lipton, C.~Elkan, B.~Narayanaswamy, Thresholding classifiers to maximize f1 score, arXiv preprint arXiv:1402.1892 (2014).

\bibitem{anguita2012k}
D.~Anguita, L.~Ghelardoni, A.~Ghio, L.~Oneto, S.~Ridella, The'k'in k-fold cross validation., in: ESANN, 2012, pp. 441--446.

\bibitem{jothimani2022mff}
S.~Jothimani, K.~Premalatha, Mff-saug: Multi feature fusion with spectrogram augmentation of speech emotion recognition using convolution neural network, Chaos, Solitons \& Fractals 162 (2022) 112512.

\bibitem{uddin2023efficacy}
M.~A. Uddin, M.~S.~U. Chowdury, M.~U. Khandaker, N.~Tamam, A.~Sulieman, The efficacy of deep learning-based mixed model for speech emotion recognition, Computers, Materials \& Continua 2023 74~(1) (2023) 1709--1722.

\bibitem{barkur2022ensemblewave}
R.~Barkur, D.~Suresh, M.~K. TN, A.~Narasimhadhan, et~al., Ensemblewave: An ensembled approach for automatic speech emotion recognition, in: 2022 IEEE International Conference on Electronics, Computing and Communication Technologies (CONECCT), IEEE, 2022, pp. 1--6.

\bibitem{dair2021linguistic}
Z.~Dair, R.~Donovan, R.~O'Reilly, Linguistic and gender variation in speech emotion recognition using spectral features, arXiv preprint arXiv:2112.09596 (2021).

\bibitem{chakraborty2022phase}
C.~Chakraborty, T.~K. Dash*, G.~Panda, S.~S. Solanki, Phase-based cepstral features for automatic speech emotion recognition of low resource indian languages, Transactions on Asian and Low-Resource Language Information Processing (2022).

\bibitem{thakur2022language}
A.~Thakur, S.~K. Dhull, Language-independent hyperparameter optimization based speech emotion recognition system, International Journal of Information Technology (2022) 1--9.

\bibitem{bhattacharya2022emotion}
S.~Bhattacharya, S.~Borah, B.~K. Mishra, A.~Mondal, Emotion detection from multilingual audio using deep analysis, Multimedia Tools and Applications 81~(28) (2022) 41309--41338.

\bibitem{al2022transformer}
B.~B. Al-onazi, M.~A. Nauman, R.~Jahangir, M.~M. Malik, E.~H. Alkhammash, A.~M. Elshewey, Transformer-based multilingual speech emotion recognition using data augmentation and feature fusion, Applied Sciences 12~(18) (2022) 9188.

\bibitem{hasan2020emotion}
H.~M. Hasan, M.~A. Islam, Emotion recognition from bengali speech using rnn modulation-based categorization, in: 2020 third international conference on smart systems and inventive technology (ICSSIT), IEEE, 2020, pp. 1131--1136.

\bibitem{meng2019speech}
H.~Meng, T.~Yan, F.~Yuan, H.~Wei, Speech emotion recognition from 3d log-mel spectrograms with deep learning network, IEEE access 7 (2019) 125868--125881.

\bibitem{wang2021novel}
X.~Wang, M.~Wang, W.~Qi, W.~Su, X.~Wang, H.~Zhou, A novel end-to-end speech emotion recognition network with stacked transformer layers, in: ICASSP 2021-2021 IEEE International Conference on Acoustics, Speech and Signal Processing (ICASSP), IEEE, 2021, pp. 6289--6293.

\end{thebibliography}





\end{document}